\documentclass[%
 aip,
 amsmath,amssymb,
 reprint,%
]{revtex4-1}

\usepackage{graphicx}
\usepackage{dcolumn}
\usepackage{bm}

\usepackage[utf8]{inputenc}
\usepackage[T1]{fontenc}
\usepackage{mathptmx}
\usepackage{xcolor}
\usepackage{mathrsfs,amsmath}
\usepackage{amsmath,esint}
\usepackage{mathtools}

\def\r{\mathbf{r}}
\def\q{\mathbf{q}}

\def\p{\mathbf{p}}

\def\w{\mathbf{w}}
\def\y{\mathbf{y}}

\begin{document}

\preprint{AIP/123-QED}

\title[Adaptive ML for CDI]{Adaptive 3D convolutional neural network-based reconstruction method for 3D coherent diffraction imaging}


\author{Alexander Scheinker}%
\email{ascheink@lanl.gov}
\author{Reeju Pokharel}
\email{reeju@lanl.gov}

\affiliation{ 
Los Alamos National Laboratory, Los Alamos, NM 87545, USA
}%

\date{\today}

\begin{abstract}
We present a novel adaptive machine-learning based approach for reconstructing three-dimensional (3D) crystals from coherent diffraction imaging (CDI). We represent the crystals using spherical harmonics (SH) and generate corresponding synthetic diffraction patterns. We utilize 3D convolutional neural networks (CNN) to learn a mapping between 3D diffraction volumes and the SH which describe the boundary of the physical volumes from which they were generated. We use the 3D CNN-predicted SH coefficients as the initial guesses which are then fine tuned using adaptive model independent feedback for improved accuracy.
\end{abstract}

\maketitle

\section{\label{sec:introduction}Introduction}
In-situ characterization of detailed 3D views of defects and interfaces and their evolution at the mesoscale (few nm - hundreds of $\mu$m) are required to develop microstructure-aware physics-based models and to design advanced materials with tailored properties \cite{ref:mesoscale2,ref:mesoscale1}. Coherent Diffraction Imaging (CDI) is a non-destructive X-ray imaging technique providing 3D measurements of sample electron density at nm resolution from which sub nm atomic displacement estimates can be calculated to understand deformation for $\mu$m sized non-crystalline specimens \cite{ref:CDI_noncrystal,ref:Au_Robinson,ref:CDI_deformation,ref:CDI_Robinson,ref:CDI1,ref:CDI_strain}. 

CDI has now been applied for a wide range of scientific studies including biology, physics and engineering \cite{ref:CDI_review}. CDI has been used to measure the 3D structures of individual viruses \cite{ref:CDI_virus} and bacteria \cite{ref:CDI_bacteria}, for imaging quantum dots \cite{ref:CDI_quantum}, to image red blood cells infected with malaria \cite{ref:CDI_blood}, for 3D imaging of human chromosomes \cite{ref:CDI_chromosome}, for imaging the 3D electron denisty of large ZnO crystals \cite{ref:CDI_ZnO}, using only partially coherent light \cite{ref:CDI_PC}, for measuring 3D lattice distortions due to defect structures in ion-implanted nano-crystals \cite{ref:BCDI:Felix1}, and for measuring dislocations in polycrystalline samples \cite{ref:CDI_RP}.

The CDI technique records only the intensity of the complex diffraction pattern originating from the illuminated sample volume, in which all phase information is lost. If the phase information in the diffraction signal could be measured, then a simple inverse Fourier transform would provide the 3D electron density which generated the diffraction pattern. Many iterative numerical methods for achieving phase retrieval have been developed which map measured diffraction patterns to electron density \cite{ref:CDI_phase_1,ref:CDI_phase_2,ref:CDI_phase_3,ref:CDI_phase_4,ref:CDI_phase_5,ref:CDI_phase_6,ref:CDI_phase_7,ref:CDI_phase_8,ref:CDI_phase_9,ref:CDI_mematic,ref:CDI_raytrace,ref:BCDI:Sid2,ref:BCDI:Sid3}. The existing CDI phase reconstruction methods are sometimes very lengthy processes requiring extensive fine tuning by expert users. Existing algorithms for inverting diffraction signals to produce a real-space image are sometimes brute-force and usually very computationally expensive. Additionally, iterative phase retrieval algorithms require expert knowledge, relying on a wide range of experience and expertise in various fields. Standard methods are sensitive to small variation in diffraction signals and different users may produce inconsistent reconstructions depending on the experience of the user and the choice of initial guesses for the parameters while expert users are able to combine various conventional algorithms such as error reduction, difference map, shrinkwrap and hybrid input-output to be capable of exploiting the available frequency information in a CDI measurement, utilizing robust treatments of measurement signal noise \cite{ref:CDInoise}. The challenges of iterative phase retrieval make it a good candidate for utilizing machine learning methods. Although ML methods cannot substitute for traditional algorithms, they do have the potential to help with the speed of obtaining reconstructions by providing an initial guess which is then fine tuned by traditional methods to achieve accurate results.

Machine learning (ML) tools, such as deep neural networks, have recently grown in popularity due to their ability to learn input-output relationships of large complex systems. Neural networks have been used to speed up lattice quantum Monte Carlo simulations \cite{ref:ML_montecarlo}, for studying complicated many body systems \cite{ref:ML_manybody}, for depth prediction in digital holography \cite{ref:ML_hologram}, and combined with model-independent feedback for adaptively controlling particle accelerator beams \cite{ref:ESML_AS}.

Recently 2D convolutional neural networks (CNN) have been utilized for speeding up diffraction-based reconstructions. CNNs have been developed to directly map 2D diffraction amplitude measurements to the amplitudes and phases of the 2D objects from which they originated, presenting an approach for orders of magnitude faster 2D amplitude and phase reconstructions for CDI \cite{ref:realtimeCDI}. CNNs have also been recently developed for orders of magnitude faster mapping of electron backscatter diffraction (EBSD) patterns to crystal orientations \cite{ref:EBSD_RP}. In this work we utilize Tensorflow and the automatic differentiation capabilities of the software package, a recent application of automatic differentiation to problem of phase retrieval is given in \cite{ref:autoDiff}.

\section{\label{sec:results}Summary of Main Results}

\begin{figure*}
\includegraphics[width=1.0\textwidth]{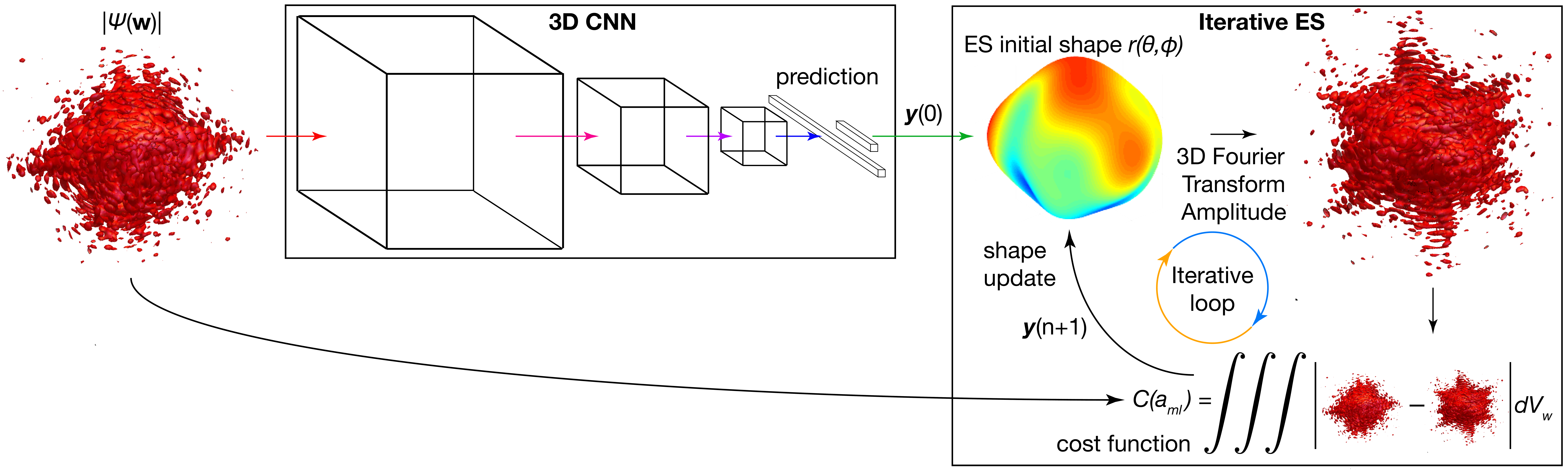}
\caption{\label{fig:ESML} The 3D CNN's output is used as the initial condition for ES tuning.}
\end{figure*}

A graphical summary of the method proposed in this work is shown in Figure \ref{fig:ESML}. We present an adaptive ML approach to the reconstruction of 3D object with uniform electron densities from synthetic diffraction patterns. Our adaptive ML framework utilizes a combination of a 3D convolutional neural network together with an ensemble of model-independent adaptive feedback agents to reconstruct 3D volumes based only on CDI diffraction measurements. The algorithm uses 3D diffracted intensities as inputs and provides outputs in the form of spherical harmonics which describe the surfaces of the 3D objects with uniform densities that generated the diffracted intensities. 

\section{\label{sec:math}Mathematical Background}

Ideally, the goal of the the CDI measurements would be to record the complex diffracted scalar wavefield $\psi(\w) = \left | \psi(\w) \right | \exp \left [ i \phi(\w) \right ]$, which is related to the Fourier transform of the electron density, $\rho(\r)$ of the sample, where $\r = (x,y,z)$ is the sample space and $\w = (w_x,w_y,w_z)$ is reciprocal space coordinates, respectively. If such a measurement could be made, the 3D electron density could be reconstructed by simply performing an inverse Fourier transform. Unfortunately, when a coherent X-ray passes through a material with electron density $\rho(\r)$, what is recorded on a detector is the intensity of the diffracted light, given by
\begin{eqnarray}
	I(\w) &=& \iint \rho(\r_1)\rho^\star(\r_2)\exp\left [ i \q \left ( \r_1 - \r_2 \right ) \right ] d\r_1d\r_2 \nonumber \\
	&=& \psi(\w)\psi^\star(\w) \nonumber \\
	&=& \left | \psi(\w) \right |^2 \exp \left [ i \phi(\w) \right ] \exp \left [ -i \phi(\w) \right ] \nonumber \\
	&=& \left | \psi(\w) \right |^2, \label{FFT}
\end{eqnarray}
with all of the phase information lost \cite{ref:CDI_Robinson,ref:Au_Robinson}. Reconstructing $\hat{\psi}(\w)$ requires lengthy phase retrieval algorithms which are typically carried out after the experiments and performed by expert users. 
\subsection{Spherical harmonics shape descriptors}
Our approach is to represent the unknown electron density inside a 3D object by a collection of basis vectors in the form of spherical harmonics, which describe the surface that encloses the volume of material of interest. Spherical harmonics are a generalization of the 1D Fourier series for representing functions defined on the unit sphere.
For any $D>0$, the Hilbert space of real square-integrable functions defined over the interval $x \in [0,D]$ is defined as
\begin{equation}
    L^2[0,D] = \left \{ f(x) \ : \ \int_{0}^{D} \left |f(x) \right|^2dx < \infty \right \}
\end{equation}
with the inner product of any $f,g \in L^2[0,D]$ defined as
\begin{equation}
    \left < f(x), g(x) \right > = \int_{0}^{D}f(x)g(x)dx. \label{inner}
\end{equation}
Distance between functions in $L^2[0,D]$ is defined by the metric
\begin{equation}
    \left \| f - g \right \|_2 = \left < f-g,f-g \right > = \int_{0}^{D}\left | f(x)-g(x)\right |^2dx.
\end{equation}
It is well known from Fourier analysis that any function $f(x) \in L^2[0,D]$ can be approximated arbitrarily closely by a linear combination of the basis functions
\begin{equation}
    \varphi_{c,n}(x) = \cos\left ( \frac{2\pi n x}{D} \right ), \quad \varphi_{s,n}(x) = \sin\left ( \frac{2\pi n x}{D} \right ), \quad n \in \mathbb{N}.
\end{equation}
If a sequence of functions $f_N$ are defined as
\begin{eqnarray}
    f_N(x) &=& c_0 + \sum_{n=1}^{N}\left [ c_n\varphi_{c,n}(x) + s_n\varphi_{s,n}(x) \right ], \\ 
    c_0 &=& \frac{1}{D}\left < f(x),1 \right > = \frac{1}{D}\int_{0}^{D}f(x)dx, \\ 
    \quad c_{n>0} &=& \frac{2}{D}\left < f(x),\varphi_{c,n}(x) \right > = \frac{2}{D}\int_{0}^{D}f(x)\varphi_{c,n}(x)dx, \\
    s_{n} &=& \frac{2}{D}\left < f(x),\varphi_{s,n}(x) \right > = \frac{2}{D}\int_{0}^{D}f(x)\varphi_{s,n}(x)dx,
\end{eqnarray}
then
\begin{equation}
    \lim_{N\rightarrow \infty} \left \| f - f_N \right \|_2 = 0.
\end{equation}

\begin{figure*}
\includegraphics[width=1.0\textwidth]{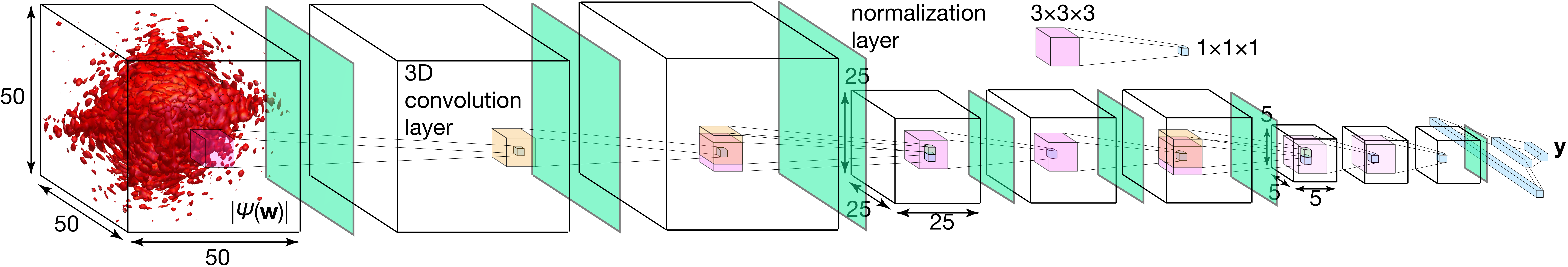}
\caption{\label{fig:3D_CNN} Overview of the 3D CNN directly using the intensity of the Fourier transform as input with a final output of dimension 28 of the coefficient of the even spherical harmonics $Y_{l}^{m}$ for $l \leq 6$: $\mathbf{y} = \left ( y_1, \dots , y_{28} \right ) =  \left ( a_{00}, a_{-22}, \dots, a_{22}, \dots, a_{66} \right )$.}
\end{figure*}

For any function $s(\theta,\phi)$ defined on the surface of the sphere, where $\theta \in [0,\pi]$ and $\phi \in [0,2\pi]$ are the spherical coordinates, the function can be approximated arbitrarily accurately with a representation of the form
\begin{equation}
    s_N(\theta,\phi) = \sum_{l=0}^{N}\sum_{m=-l}^{l}a_{ml}Y^m_l(\theta,\phi), \label{rhat}
\end{equation}
where the coefficients are found by the inner product
\begin{eqnarray}
    a_{ml} &=& \left < s(\theta,\phi), Y_{l}^{m}(\theta,\phi) \right> \nonumber \\
    &=& \int_{0}^\pi \int_{0}^{2\pi} r(\theta,\phi){Y_{l}^{m}}^{*}(\theta,\phi) \sin(\theta) d\theta d\phi.
\end{eqnarray}

By approximating in terms of the basis of spherical harmonics, we assume that we can find a star-convex approximation of a surface $s(\theta,\phi)$.  

\section{\label{sec:ES}Adaptive Machine Learning for Phase Retrieval}
 To determine the unknown electron density $\rho(\r,\theta,\phi)$, we make the assumption that the electron density is non-zero only within some compact set and that the density is uniform within some bounding surface $\partial \rho(\r,\theta,\phi) = s(\theta,\phi)$ of the form
\begin{equation}
    \rho(\r,\theta,\phi) =
    \begin{cases}
      d, & |\r| \leq s(\theta,\phi) \\
      0,        & |\r| \geq s(\theta,\phi)
    \end{cases}. \label{case}
\end{equation}
Note that in this proof-of-concept work, we are considering solid objects of uniform density $d$ without internal structures. Our 3D reconstruction approach is to find a set of coefficients $\hat{a}_{ml}$ up to order $l=N$, $\y=\left (\hat{a}_{00},\dots,\hat{a}_{ml},\dots,\hat{a}_{NN} \right )$, which define a surface that approximates $s(\theta,\phi)$ by constructing
\begin{equation}
    \hat{s}(\theta,\phi) = \sum_{l=0}^{N}\sum_{m=-l}^{l}\hat{a}_{ml}Y^m_l(\theta,\phi), \label{hatshat}
\end{equation}
which in turn defines an electron density
\begin{equation}
    \hat{\rho}(\r,\theta,\phi) =
    \begin{cases}
      d, & |\r| \leq \hat{s}(\theta,\phi) \\
      0,        & |\r| \geq \hat{s}(\theta,\phi)
    \end{cases}, \label{case_2}
\end{equation}
that approximates $\rho(\r,\theta,\phi)$. In order to find the appropriate spherical harmonics, we calculate the amplitude of the 3D Fourier transform $\left | \mathcal{F}\left (\hat{\rho}(\r,\theta,\phi) \right ) \right |$ which represents the amplitude of a complex scalar diffracted wavefield $\left | \hat{\psi}(\w) \right |$ and then compare it to the ground-truth synthetic 3D diffraction pattern.

\subsection{3D Convolutional Neural Network}

Our approach uses a combination of a 3D convolutional neural network together with model-independent adaptive feedback. Convolutional neural networks are very powerful tools that can learn relationships between parameters in complex systems and in this case can directly utilize spatial information to learn 3D features from the 3D amplitude of the Fourier transform. The architecture of the 3D CNN network developed for our problem is shown in Figure (\ref{fig:3D_CNN}).

We point out that in mapping 3D Fourier transform intensities to spherical harmonic coefficients, a CNN is only able to predict the even $l$-valued harmonics $Y_{0}^{m},Y_{2}^{m},Y_{4}^{m},\dots$ for the following reason. 

\begin{figure*}
\includegraphics[width=1.0\textwidth]{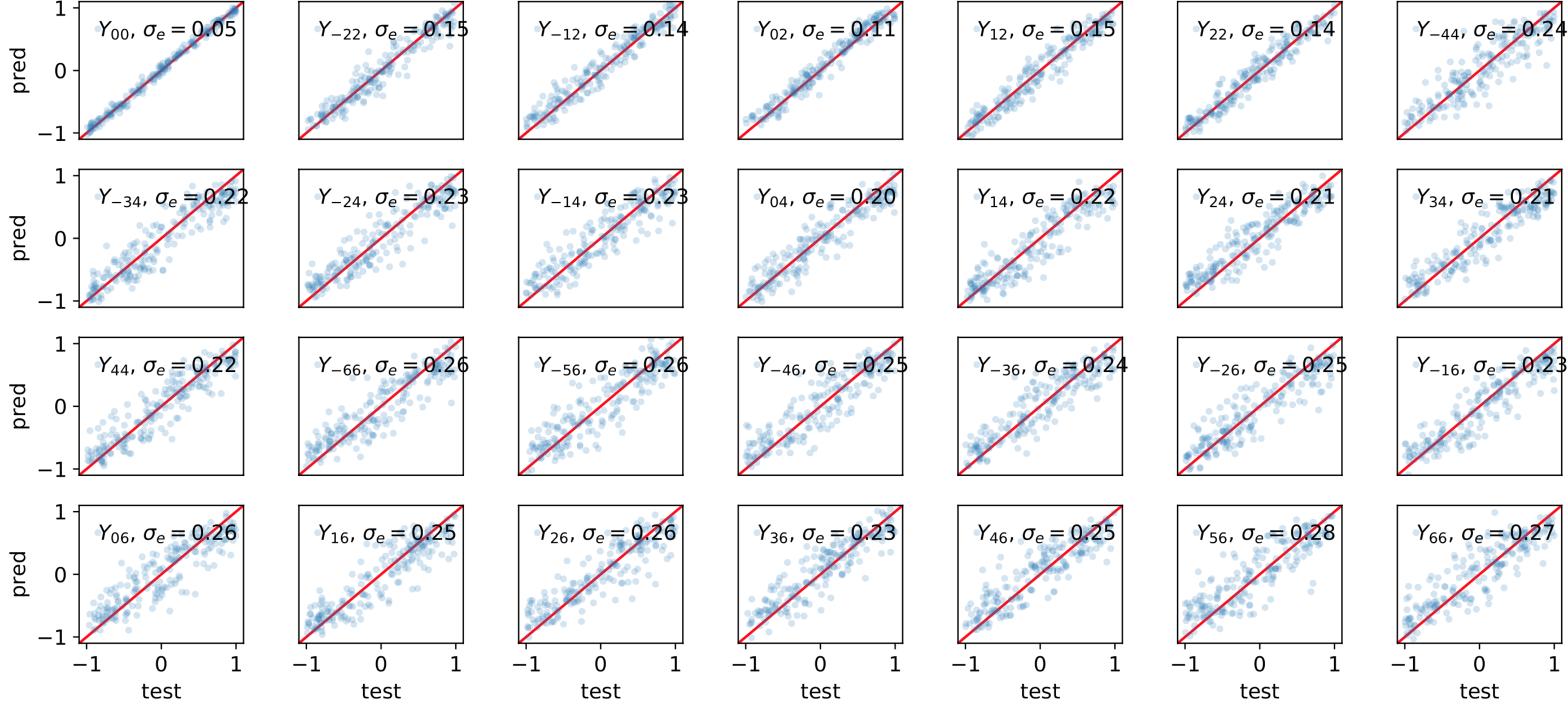}
\caption{\label{fig:3D_CNN_xy} Test vs prediction values shown for the even-valued $a_{lm}$ coefficients for 200 test structures along with the standard deviation of the error for each coefficient.}
\end{figure*}

Consider two volumes described by surfaces which are perturbations of a sphere, of the form
\begin{equation}
    s_{\pm}(\theta,\phi) = Y_{0}^{0}(\theta,\phi) \pm \epsilon Y_{l}^{m}(\theta,\phi), \label{s_odd}
\end{equation}
where $l$ is odd. The odd $l$-valued harmonics are themselves odd functions because all real spherical harmonics satisfy:
\begin{eqnarray}
    (-1)^lY_{l}^{m}(\theta,\phi) &=& Y_{l}^{m}(\pi - \theta,\pi + \phi) \nonumber \\
    &\Longrightarrow& -Y_{l_{\mathrm{odd}}}^{m}(\theta,\phi) = Y_{l_{\mathrm{odd}}}^{m}(\pi - \theta,\pi + \phi). \nonumber
\end{eqnarray}
Therefore the two surfaces in (\ref{s_odd}) can be rewritten as
\begin{eqnarray}
    s_{+}(\theta,\phi) &=& Y_{0}^{0}(\theta,\phi) + \epsilon Y_{l}^{m}(\theta,\phi), \nonumber \\
    s_{-}(\theta,\phi) &=& Y_{0}^{0}(\theta,\phi) + \epsilon Y_{l}^{m}(\pi - \theta,\pi + \phi), \label{s_odd2}
\end{eqnarray}
which are simply reflections and so the intensities of the Fourier transforms of their volumes $\rho_{\pm}$ are indistinguishable because of the lost phase information. When teaching a neural network to map diffraction patterns to coefficients, we end up giving it inputs generated from two different surfaces and volumes, with their corresponding spherical harmonic coefficients as the correct outputs:
\begin{eqnarray}
    s_{+} &\Longrightarrow& \rho_+ \Longrightarrow \left | \mathcal{F}\left (\rho_+ \right ) \right | = \left | \mathcal{F}\left (\rho_\pm \right ) \right | \Longrightarrow \mathrm{CNN} \Longrightarrow \{1,\epsilon\}, \nonumber \\
    s_{-} &\Longrightarrow& \rho_- \Longrightarrow \left | \mathcal{F}\left (\rho_- \right ) \right | = \left | \mathcal{F}\left (\rho_\pm \right ) \right | \Longrightarrow \mathrm{CNN} \Longrightarrow \{1,-\epsilon\}. \nonumber
\end{eqnarray}
Because in this case the Fourier intensities are exactly the same after learning over thousands of random data sets, the neural network is confused and at best can predict only the average value of $0$ for all of the odd spherical harmonic coefficients.

This problem can be confirmed numerically in three ways: 1). If only positive values of odd harmonics are used to generate volumes then the CNN learns how to map diffraction patters to both even and odd harmonics, but this limits its applicability because realistic objects have shapes that are composed of both odd and even harmonic components. 2). If the network is tasked with only identifying the magnitude of the odd harmonics it is able to learn the relationship, but resulting structure predictions must then iterate through all of the possible $\pm$ combinations to find those which best match the given Fourier transform intensity to calculate the correct object shape, but the created object's orientation will not necessarily match that of the target. 3). Finally, if the CNN is given the actual 3D electron densities or the 3D Fourier transforms (not just intensities) which contain the phase information as inputs, then it learns to map all spherical harmonics correctly, but this is not helpful for our problem, where the goal is to make predictions solely based on diffracted intensities.

This limitation of a neural network approach was also documented in \cite{ref:realtimeCDI} where they found that the neural network would sometimes predict objects that "are twin images of each other, and that they can be obtained from each other through a centrosymmetric inversion and complex conjugate operation. Both images are equivalent solutions to the input diffraction pattern." Because of the limitations described above, the CNN was trained to map 3D diffracted intensities to only the even valued spherical harmonic coefficients that describe the boundaries of the volumes whose Fourier transforms generated those intensities.  

Data for training the 3D CNN was generated by sampling coefficients from uniform distributions with ranges:
\begin{equation}
    \hat{a}_{ml} \in \left[-c_{ml}, c_{ml} \right ], \qquad c_{ml} = \frac{0.25}{1+l+|m|}, \nonumber
\end{equation}
and generating training volumes based on surfaces of the form
\begin{equation}
    \hat{s}(\theta,\phi) = \frac{3}{2}Y_{0}^{0}(\theta,\phi) + \sum_{l=0}^{N=6}\sum_{m=-l}^{l}\hat{a}_{ml}Y^m_l(\theta,\phi),
\end{equation}
where the large $Y_0^0(\theta,\phi)$ value ensured that we are working with a well defined surface perturbed by higher order components similar to naturally occurring complex grain shapes.

We generated 500,000 training sets of 49 coefficients, for $l=0,\dots,N=6$, each of which was used to generate a surface and a volume bound by that surface to perform a 3D Fourier transform. The input to the CNN was the intensity of the 3D Fourier transform and the output of the CNN was a 28 dimensional vector which was compared to the 28 even spherical harmonic coefficients $Y_{l}^{m}$ for $l \leq 6$: $\mathbf{y} = \left ( y_1, \dots , y_{28} \right ) =  \left ( \hat{a}_{00}, \hat{a}_{-22}, \dots, \hat{a}_{22}, \dots, \hat{a}_{66} \right )$ via the cost function:
\begin{equation}
    C_{\mathrm{CNN}}(\y) = \sum_{j=1}^{28} \left | y_j - a_{ml} \right | = \sum_{l_{\mathrm{odd}}=0}^{6}\sum_{m=-l}^l \left |\hat{a}_{ml} - a_{ml}  \right |.
\end{equation}

\begin{figure*}
\includegraphics[width=1.0\textwidth]{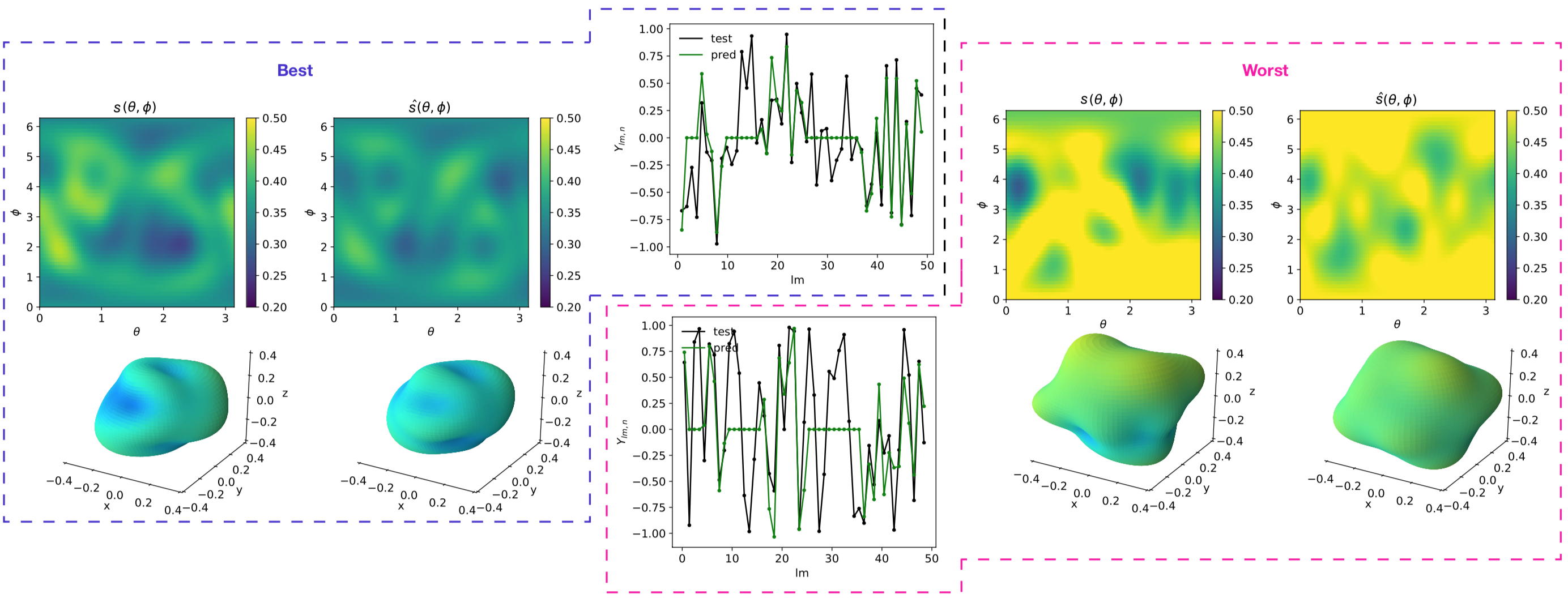}
\caption{\label{fig:3D_CNN_end} Detailed view of the best and worst performers out of 200 test structures that had not been seen during training of the CNN.}
\end{figure*}

CNN performance is illustrated in Figures \ref{fig:3D_CNN_xy} and \ref{fig:3D_CNN_end} where the predictive accuracy of 200 unseen test sets is shown along with the lowest and highest prediction accuracy shapes.

In this setup CNN performance is accurate, but can only make predictions for the even spherical harmonic coefficients. Furthermore, for use in experiments, the accuracy of a learned model-based approach such as a CNN may suffer depending on experimental setup changes and may require very lengthy experiment-specific retraining.

In order to predict the even spherical harmonics and also to make these results more robust to a wide range of experimental conditions, the next step of this approach is to use a model-independent algorithm that adjusts all spherical harmonic coefficients directly based on matching individual 3D Fourier transforms. The model independent approach is also capable of handling very large numbers of coefficients as shown below when tuning up to 225 spherical harmonics simultaneously.

\subsection{Model-independent tuning}

For the adaptive part of this work, we utilize a model-independent extremum seeking (ES) algorithm which has was originally developed for control and optimization of uncertain and time-varying systems by simultaneously tuning large numbers of coupled parameters based only on noisy measurements \cite{ref:ES_Bounded}. This bounded form of ES has been analytically studied with convergence proofs for general non-differentiable dithers \cite{ref:ES_nonC2}, has been proven to converge to optimal controllers for unknown systems \cite{ref:ES_opt}, and has been applied to automatically control charged particle beams in particle accelerators \cite{ref:ESML_AS}.

The ES method is applicable to n-dimensional dynamic system of the form
\begin{eqnarray}
	\frac{d\y}{dt} &=& f(\y,\p,t), \label{dynamic} \\
	\hat{C} &=& C(\y,t) + n(t), \label{yofx}
\end{eqnarray}
where $\y=(y_1,\dots,y_n)$ are physical quantities of interest, such as diffraction patterns of electron densities. The $\p=(p_1,\dots,p_m)$ are controlled parameters, such as the spherical harmonics that define the surface of a volume and $t$ is time. The function $f$ may be an unknown function governing the system's dynamics. $\hat{C}$ is a measurement of an analytically unknown function $C(\y,t)$ that is noise-corrupted by an unknown function of time, $n(t)$, and depends on both the parameter values $\y$ and on time due to a time-varying system environment. In our approach, we compare the intensities of the measured diffraction and generated diffraction wavefields and quantify the difference with the numerical cost function whose minimization is our goal:
\begin{equation}
    C(\y) = \frac{1}{\mu(V_\w)}\iiint_{V_\w}  \bigl\lvert \left | \psi(\w) \right | - \left | \hat{\psi}(\w) \right |  \bigl\rvert dV_\w, \label{cost}
\end{equation}
where integration is performed over a volume in reciprocal space $V_\w$ of measure $\mu(V_\w) = (w_{\mathrm{max}} - w_{\mathrm{min}})^3$.

In experimental applications of such a method, uncertainty would come from the unknown electron densities that we are trying to find and from uncertainties (such as misalignment of components and drifts in X-ray coherence volume, wavelength, and flux) in the experimental setup.

The parameters that we tuned were the $Y^l_m$ coefficients
\begin{equation}
    \y= \left (y_1,\dots,y_j,\dots,y_{(1+N)^2} \right ) = \left (\hat{a}_{00},\dots,\hat{a}_{ml},\dots,\hat{a}_{NN} \right ), \label{y_aml}
\end{equation}
which define the boundary surface of an unknown volume as in Equation (\ref{rhat}) and the function that we are minimizing is $C(\y)$ as defined in (\ref{cost}). The ES algorithm perturbs parameters according to the dynamics
\begin{equation}
    \frac{dy_j}{dt} = \sqrt{2\alpha\omega_j}\cos\left (\omega_j t + k \hat{C}(\y,t) \right ), \label{ES}
\end{equation}
where $\omega_j = \omega r_j$ and $r_i \neq r_j$ for $i\neq j$. In (\ref{ES}) $\alpha$ is a dithering amplitude which can be increased to escape local minima. Once the dynamics have settled near an equilibrium point of (\ref{ES}), which may be a local minimum of $C$, each parameter will continue to oscillate about its local optimal value with a magnitude of $\sqrt{2\alpha/\omega_j}$. The term $k>0$ is a feedback gain. For $\omega \gg 1$ the dynamics (\ref{ES}) are on average approximated by
\begin{equation}
    \frac{d\y}{dt} = -k\alpha\nabla_{\y}C(\y,t), \label{ES_ave}
\end{equation}
which tracks the time-varying minimum of the unknown function $C(\y,t)$ with respect to $\y(t)$ although using only its noise-corrupted measurement $\hat{C}$ as input. The reason behind convergence is that the evolution of the coupled parameters $y_j$ is decoupled and made orthogonal relative to the inner product in the $L^2[0,t]$ Hilbert space as defined in (\ref{inner})
\begin{equation}
    \lim_{\omega_i,\omega_j \rightarrow \infty}\left < \cos(\omega_i t),\cos(\omega_j t) \right > = 0.
\end{equation}
Details and analytical proofs are available in \cite{ref:ES_Bounded,ref:ES_opt,ref:ES_nonC2}.

\begin{figure*}
\includegraphics[width=1.0\textwidth]{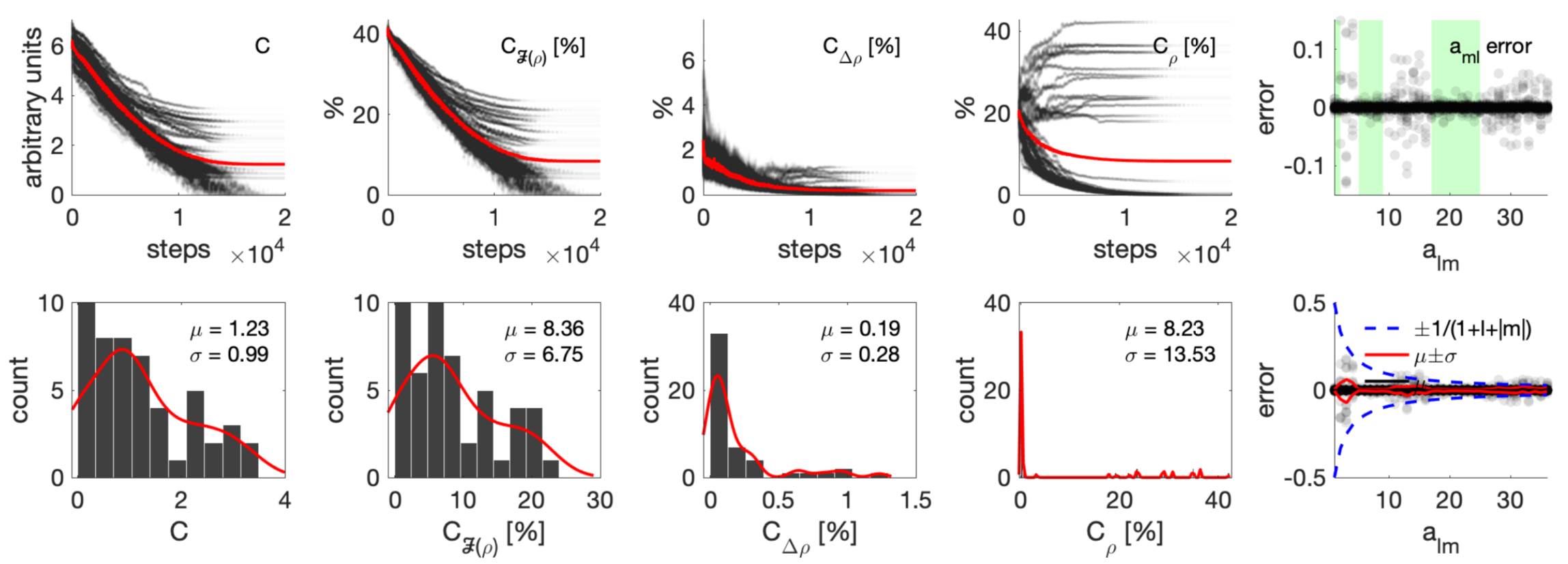}
\caption{\label{fig:ESspeed} The top images of the first four columns show the convergence of the cost function $C$ as defined in (\ref{cost}) together with convergence of the quantities defined in (\ref{Cp1})-(\ref{Cp3}). The bottom images of the first four columns show histograms of converged values after the final iteration. The top image of the last column shows the errors of the converged $\hat{a}_{ml}$ coefficients with the green band highlighting the even coefficients relative to which there is no ambiguity in the intensity of the Fourier transform. The bottom image of the last column shows the $a_{ml}$ errors together with mean and standard deviation as well as the bounds of the random distributions from which they were generated (blue/dashed).}
\end{figure*}

For iterative optimization as done in this work, we replace the continuous time dynamics (\ref{ES}) with their discrete-time approximation and make iterative updates according to
\begin{equation}
    y_j(n+1) = y_j(n) + \Delta_t \sqrt{2\alpha\omega_j}\cos \left (\omega_j \Delta_t n + k \hat{C}(n) \right ), \label{ES_it}
\end{equation}
which is a finite difference approximation of (\ref{ES}) for $\Delta_t \ll 1$.

In order to test the convergence properties of the ES algorithm, we simultaneously measured the following three quantities during convergence for 100 random volumes:
\begin{eqnarray}
    C_{\mathcal{F}(\rho)} &=& \frac{100}{\mu(V_\w)} \times \iiint_{V_\w}  \bigl\lvert \left | \psi(\w) \right | - \left | \hat{\psi}(\w) \right |  \bigl\rvert dV_\w, \label{cost} \label{Cp1} \\
    C_{\rho} &=& \frac{100}{\mu(\rho)} \times\iiint_{V} \left |\hat{\rho}(\r,\theta,\phi) - \rho(\r,\theta,\phi)  \right |dV , \label{Cp2} \\
    C_{\Delta \rho} &=& 100 \times \left |\mu(\hat{\rho}) - \mu(\rho) \right |, \label{Cp3} \\
    \mu(V_\w) &=& \iiint_{V_\w}\left | \psi(\w) \right |dV_\w, \quad V_\w = \left [ w_{\mathrm{min}},w_{\mathrm{max}}\right ]^3, \label{spec_measure} \\
    \mu(\rho) &=& \iiint_{V}\rho(\r,\theta,\phi)dV. \label{vol_measure}
\end{eqnarray}

The quantity $C_{\mathcal{F}(\rho)}$ is a measure of the percent difference between the intensity of the target and reconstructed Fourier transforms. Convergence would mean we have matched the intensity of the Fourier transforms. However, it does not guarantee the correct shape due to missing phase information, and the same 3D object could be rotated or reflected. The quantity $C_{\rho}$ is a measure of the mismatch between volumes which is non-zero when the objects have the same shape, but are of different orientations. Finally, the quantity $C_{\Delta \rho}$ is a measure of shape convergence which subtracts the total volumes occupied by the two shapes and therefore will converge to zero when the two shapes are the same even if they have different orientations.

\begin{figure*}
\includegraphics[width=1.0\textwidth]{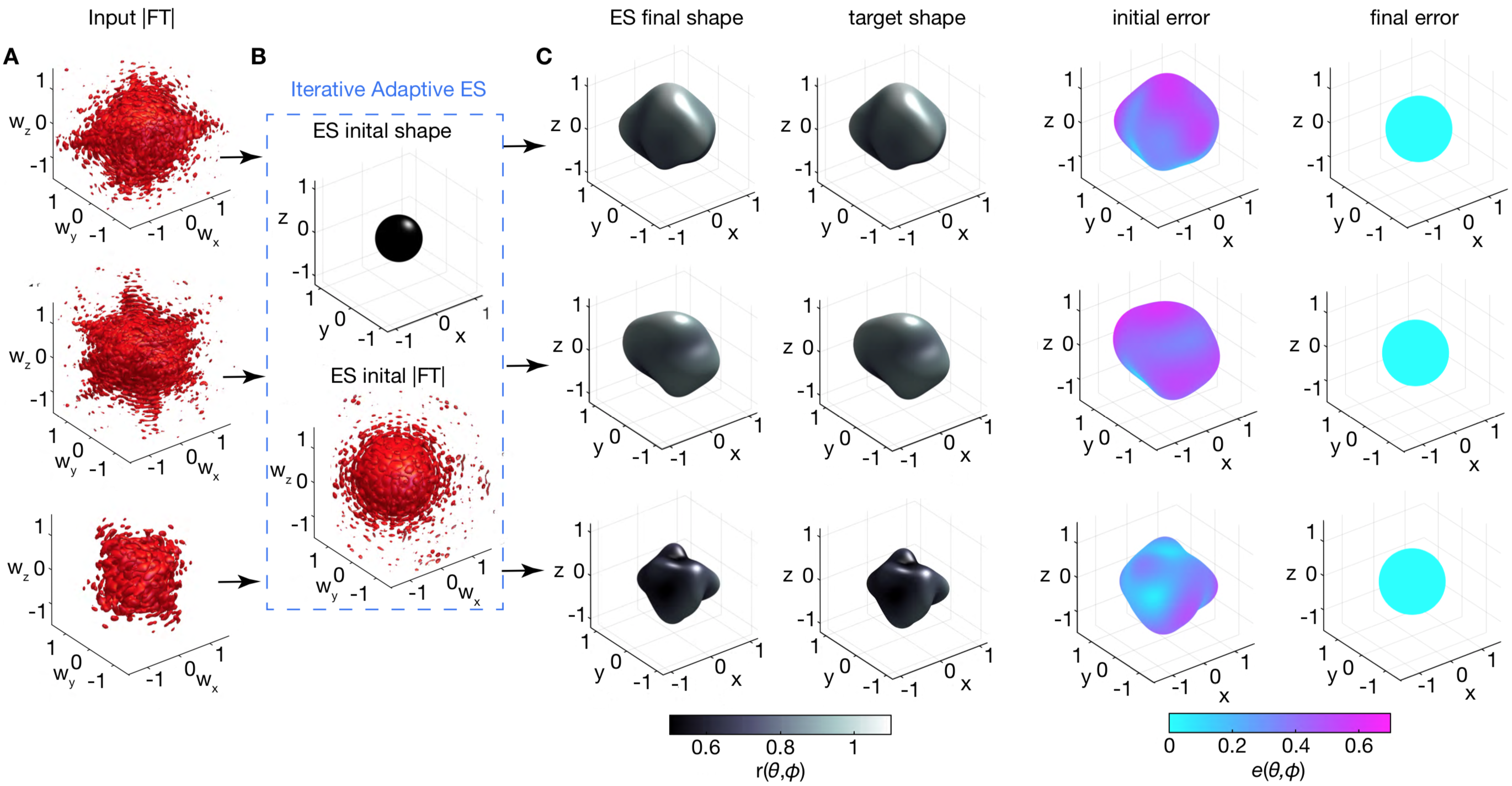}
\caption{\label{fig:ESconvergence} Convergence of the ES algorithm for 3 different structures A, B, and C. In this demonstration the algorithm always started with only $Y_{00} \neq 0$ which created an initial spherical shape as shown on the left. The reconstructed volumes perfectly matched the targets in our $50\times50\times50$ volume representations as seen by the error $r(\theta,\phi)$ showing the initial and final differences between the volume bounding surfaces mapped onto the sphere. The last column shows the convergence of all parameter errors as the algorithm is able to find the correct values for all 36 $Y_{lm}$ settings in each case.}
\end{figure*}

We created 100 random 3D shapes, generated their 3D Fourier transforms, and fed the intensities of those transforms into the 3D CNN. The predictions of the CNN were then used as the starting point for the ES algorithm. Results of ES convergence for 100 random 3D shapes are shown in Figures \ref{fig:ESspeed} and \ref{fig:ESconvergence}. Looking at the top images in the second and third columns of Figure \ref{fig:ESspeed} it is clear that the CNN-based objects had Fourier transform intensity errors, $C_{\mathcal{F}(\rho)}$ of 40$\%$ relative to their full spectrum measures as defined in (\ref{spec_measure}) and volume errors, $C_{\Delta\rho}$ of approximately 3$\%$. The bottom images of the second and third columns show that by the end of convergence the average intensity error was 8.4$\%$ and volumetric error was 0.19$\%$.

 In Figure \ref{fig:ESspeed}, it is evident that on average all of the quantities $C$, $C_{\mathcal{F}(\rho)}$, $C_{\Delta \rho}$, and $C_{\rho}$ converge towards zero; however $C_{\rho}$ has several large outliers that never converge which implies that the densities being created are of the correct shape, but wrong orientation. The last column of Figure \ref{fig:ESspeed} is showing the errors between predicted $\hat{a}_{ml}$ and correct $a_{ml}$ values. The green background in Figure \ref{fig:ESspeed} highlights the even valued coefficients which we expect to match exactly while the odd components are expected to sometimes not converge due to the ambiguity introduced by the lack of phase information in the Fourier transform's intensity, as discussed above. Overall, the results of Figure \ref{fig:ESspeed} confirm that the ES approach is very robust and is able to find the correct object shape with the possibility of an incorrect orientation in space, as expected. Figure \ref{fig:ESconvergence} shows three examples of exact agreement between 3D test objects and their ES-based reconstructions.

\subsection{\label{sec:ES_CNN}Adaptive ML for experimental data}


In order to further demonstrate the robustness of the adaptive ML approach, we applied it to an experimentally measured 3D crystal volume that was obtained using high energy diffraction microscopy (HEDM)\cite{ref:HEDM_4}. HEDM is used for non-destructive measurements of spatially resolved orientation ($\sim$ 1.5 $\mu$m and 0.01$^\circ$), grain resolved orientation, and elastic strain tensor ( 10$^{-3}$) from representative volume elements with hundreds of bulk grains in the measured microstructure (mm$^3$) \cite{ref:HEDM_1}. HEDM measurements at multiple states of a sample's evolution can be used as inputs to inform and validate crystal plasticity models \cite{ref:HEDM_2, ref:HEDM_3}. For a broad overview of HEDM and its many applications the reader is refereed to \cite{ref:HEDM_1} and the multiple references within.

\begin{figure*}
\includegraphics[width=0.9\textwidth]{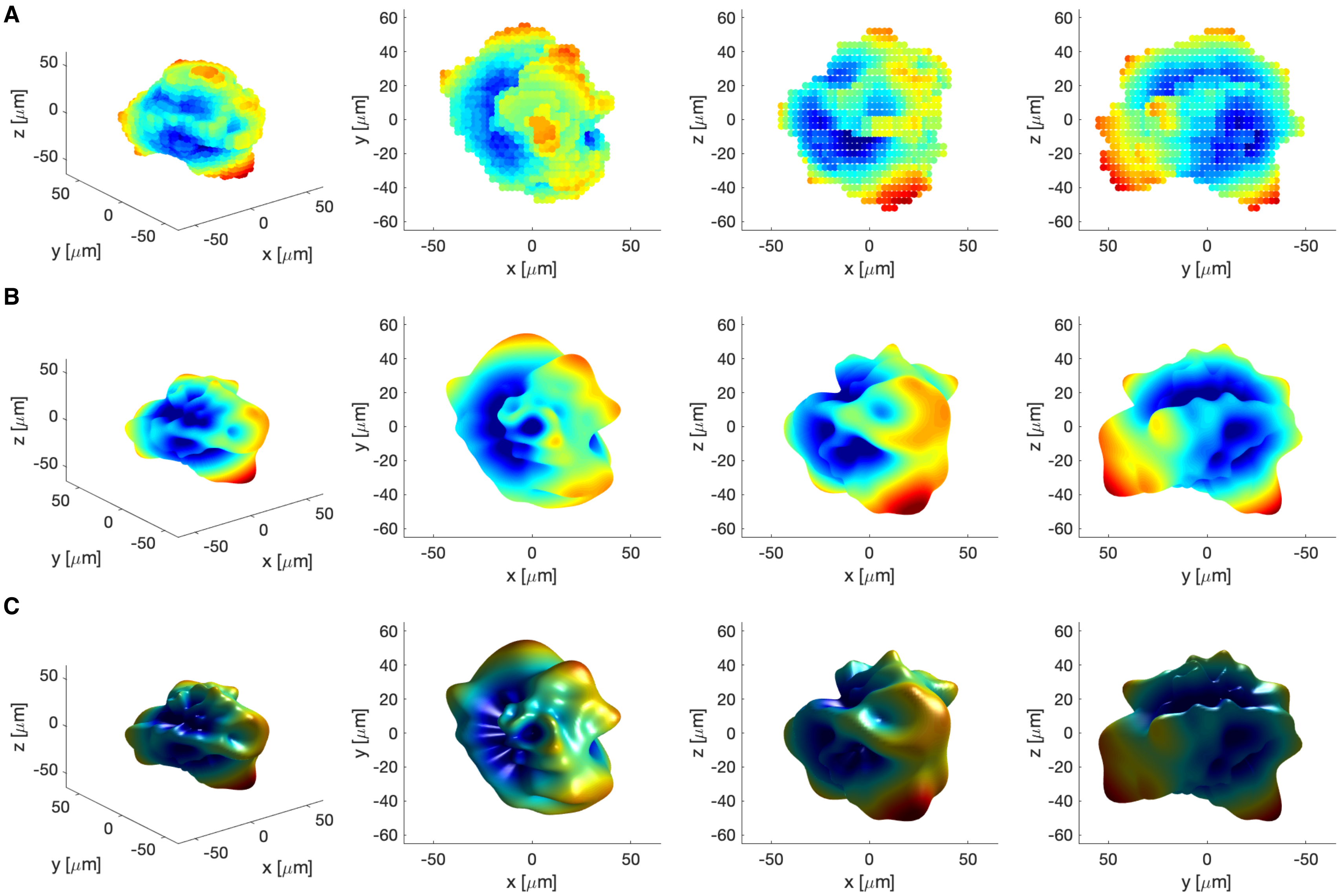}
\caption{\label{fig:g4s1_3D} Result of using the 3D CNN output as the initial guess for first the 49 $a_{lm}$ coefficients $(a_{00},a_{-11},a_{01},a_{11},\dots,a_{66})$, followed by ES fine tuning of all 225 coefficients $(a_{00},\dots,a_{1414})$. The top row (A) shows the first measured state of the HEDM structure from various views. The second row (B) shows the CNN-ES convergence results. The third row (C) is showing the same as (B) with shading for easier 3D visualization.}
\end{figure*}

\begin{figure*}
\includegraphics[width=1.0\textwidth]{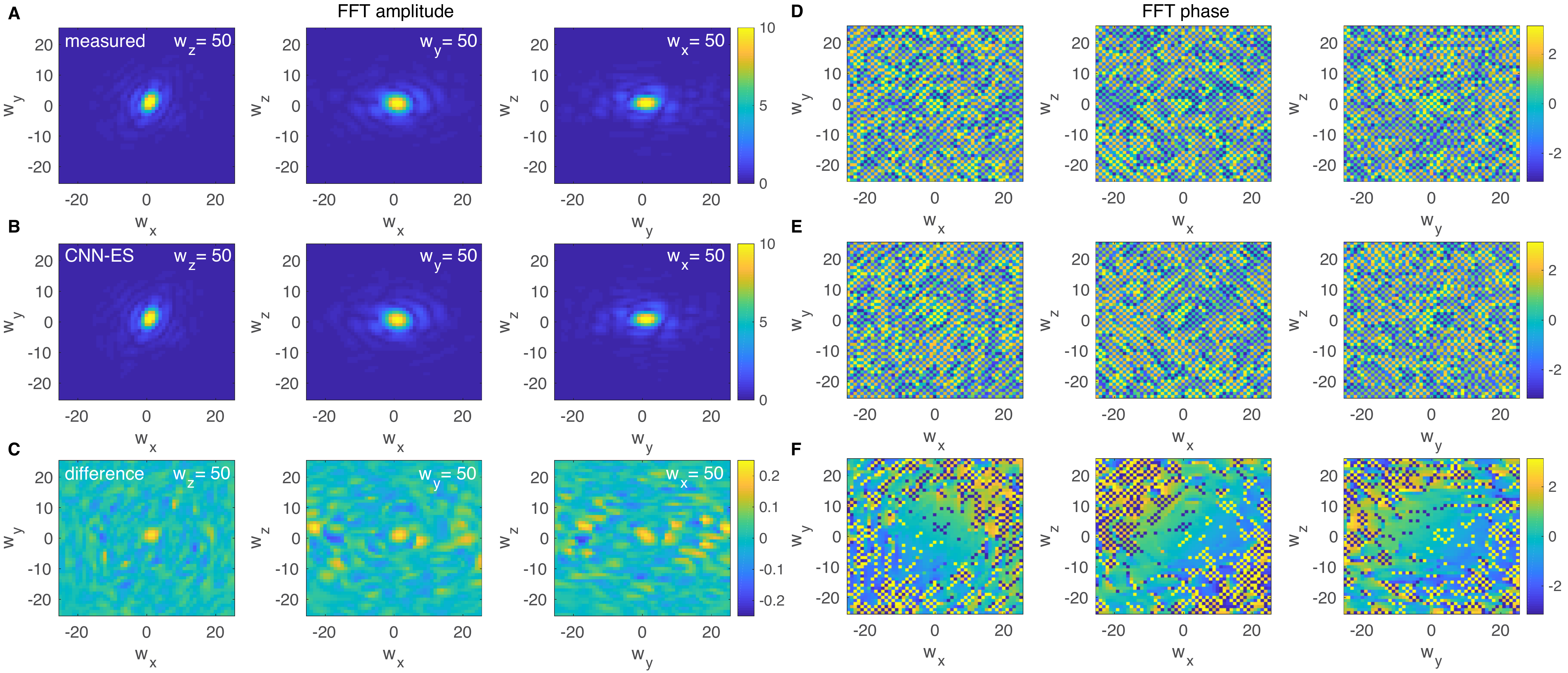}
\caption{\label{fig:g4s1_FFT} Orthogonal slices through the 3D amplitude and 3D phase of the FT are shown. The top left row (A) shows the amplitudes of the HEDM measurement. The middle left row (B) shows the amplitudes of the CNN-ES reconstruction. The bottom left row (C) shows the difference between (A) and (B), note the reduced color scale range. The rows (D), (E), and (F) show the same for the FT phase.}
\end{figure*}

To test the robustness of our adaptive ML approach to a structure that had never been seen by neither the CNN nor the ES algorithm during its tuning and design, we picked out a single 3D grain from a polycrystalline copper sample which was measured with the HEDM technique at the Advanced Photon Source (APS) \cite{ref:HEDM_4}. The intensity of the 3D Fourier transform of this volume was fed into the CNN which provided an estimate of the first 28 even $a_{ml}$ coefficients. These were then fed as initial guesses into the ES adaptive feedback algorithm which had the freedom to tune all 225 coefficients of the $l \in \{ 0,\dots,14 \}$ $Y^m_l(\theta,\phi)$ spherical harmonics in order to match the generated and measured diffraction patterns. The 3D shape and 2D slices of the amplitude and phase of the reconstructed particle results of convergence are shown in Figures \ref{fig:g4s1_3D} and \ref{fig:g4s1_FFT}.

The HEDM grain is relatively large with $\sim$60 $\mu$m diameter and is therefore too large to be imaged with existing CDI techniques due to light energy and coherence length limitations of existing light sources. Nevertheless, for testing the proposed method, the morphology of the HEDM crystal was interesting in its complexity and was similar to what has been measured by Bragg CDI techniques as applied to quantum dot nanoparticles \cite{ref:CDI_review}. Furthermore, advanced light sources such as the planned Linac Coherent Light Source II (LCLS-II) free electron laser (FEL) and the APS Upgrade (APSU) are expected to have increased transverse and longitudinal coherence lengths with techniques such as self seeding \cite{ref:LCLS,ref:APSU}, to image larger than 1 $\mu$m diameter crystals using high-energy CDI combined with HEDM \cite{ref:BCDI:Sid1}. 

\section{\label{sec:conclusions}Conclusions}

In this proof-of-concept work, we demonstrate reconstructions of arbitrary 3D shapes, while assuming no contribution of internal lattice distortions to the diffraction signal. One immediate limitation of this method is that by parameterizing surfaces as single valued functions over the 2D domain $(\theta,\phi) \in [0,\pi]\times[0,2\pi]$, we are limiting ourselves to producing only star convex shapes. Star convex shapes are ones in which a line can be drawn from the center point to the outer edge without intercepting any other edges and therefore do not include more complex surfaces which are not simply connected, such as a donut-like shape with holes. Generalization of this approach to a larger class of shapes will be the study of future work by utilizing surface parameterization which decomposes a 3D particle surface onto three orthogonal directions \cite{ref:Sph_1,ref:Sph_2,ref:Sph_3}.

Furthermore, the method presented here can be readily extended to reconstruct additional phases for crystals with internal structures due to inherent defects and dislocations by several methods including the use of generative convolutional neural networks or by extending the adaptive model-independent process to include more degrees of freedom. Although the CNN model is trained only on synthetic diffraction data, the adaptive framework will readily account for noise in the experimental data for robust reconstruction of 3D crystals with internal structures, which is a topic of future work.

\begin{acknowledgments}
This work was supported by the US Department of Energy through the Los Alamos National Laboratory. Los Alamos National Laboratory is operated by Triad National Security, LLC, for the National Nuclear Security Administration of U.S Department of Energy (Contract No. 89233218CNA000001). We acknowledge the support provided by the Institute of Materials Science Rapid Response project RR2020-R\&D-1. 
\end{acknowledgments}

\section{Data Availability} The data that support the findings of this study are available from the corresponding authors upon reasonable request.

\nocite{*}
\bibliography{3DCNNES_8_23_20}

\providecommand{\noopsort}[1]{}\providecommand{\singleletter}[1]{#1}%
\begin{thebibliography}{52}%
\makeatletter
\providecommand \@ifxundefined [1]{%
 \@ifx{#1\undefined}
}%
\providecommand \@ifnum [1]{%
 \ifnum #1\expandafter \@firstoftwo
 \else \expandafter \@secondoftwo
 \fi
}%
\providecommand \@ifx [1]{%
 \ifx #1\expandafter \@firstoftwo
 \else \expandafter \@secondoftwo
 \fi
}%
\providecommand \natexlab [1]{#1}%
\providecommand \enquote  [1]{``#1''}%
\providecommand \bibnamefont  [1]{#1}%
\providecommand \bibfnamefont [1]{#1}%
\providecommand \citenamefont [1]{#1}%
\providecommand \href@noop [0]{\@secondoftwo}%
\providecommand \href [0]{\begingroup \@sanitize@url \@href}%
\providecommand \@href[1]{\@@startlink{#1}\@@href}%
\providecommand \@@href[1]{\endgroup#1\@@endlink}%
\providecommand \@sanitize@url [0]{\catcode `\\12\catcode `\$12\catcode
  `\&12\catcode `\#12\catcode `\^12\catcode `\_12\catcode `\%12\relax}%
\providecommand \@@startlink[1]{}%
\providecommand \@@endlink[0]{}%
\providecommand \url  [0]{\begingroup\@sanitize@url \@url }%
\providecommand \@url [1]{\endgroup\@href {#1}{\urlprefix }}%
\providecommand \urlprefix  [0]{URL }%
\providecommand \Eprint [0]{\href }%
\providecommand \doibase [0]{http://dx.doi.org/}%
\providecommand \selectlanguage [0]{\@gobble}%
\providecommand \bibinfo  [0]{\@secondoftwo}%
\providecommand \bibfield  [0]{\@secondoftwo}%
\providecommand \translation [1]{[#1]}%
\providecommand \BibitemOpen [0]{}%
\providecommand \bibitemStop [0]{}%
\providecommand \bibitemNoStop [0]{.\EOS\space}%
\providecommand \EOS [0]{\spacefactor3000\relax}%
\providecommand \BibitemShut  [1]{\csname bibitem#1\endcsname}%
\let\auto@bib@innerbib\@empty
\bibitem [{\citenamefont {McDowell}(2010)}]{ref:mesoscale2}%
  \BibitemOpen
  \bibfield  {author} {\bibinfo {author} {\bibfnamefont {D.~L.}\ \bibnamefont
  {McDowell}},\ }\bibfield  {title} {\enquote {\bibinfo {title} {A perspective
  on trends in multiscale plasticity},}\ }\href {\doibase
  https://doi.org/10.1016/j.ijplas.2010.02.008} {\bibfield  {journal} {\bibinfo
   {journal} {International Journal of Plasticity}\ }\textbf {\bibinfo {volume}
  {26}},\ \bibinfo {pages} {1280--1309} (\bibinfo {year} {2010})}\BibitemShut
  {NoStop}%
\bibitem [{\citenamefont {Crabtree}\ and\ \citenamefont
  {Sarrao}(2012)}]{ref:mesoscale1}%
  \BibitemOpen
  \bibfield  {author} {\bibinfo {author} {\bibfnamefont {G.}~\bibnamefont
  {Crabtree}}\ and\ \bibinfo {author} {\bibfnamefont {J.}~\bibnamefont
  {Sarrao}},\ }\bibfield  {title} {\enquote {\bibinfo {title} {Opportunities
  for mesoscale science},}\ }\href {\doibase
  https://doi.org/10.1557/mrs.2012.274} {\bibfield  {journal} {\bibinfo
  {journal} {MRS bulletin}\ }\textbf {\bibinfo {volume} {37}},\ \bibinfo
  {pages} {1079--1088} (\bibinfo {year} {2012})}\BibitemShut {NoStop}%
\bibitem [{\citenamefont {Miao}\ \emph {et~al.}(1999)\citenamefont {Miao},
  \citenamefont {Charalambous}, \citenamefont {Kirz},\ and\ \citenamefont
  {Sayre}}]{ref:CDI_noncrystal}%
  \BibitemOpen
  \bibfield  {author} {\bibinfo {author} {\bibfnamefont {J.}~\bibnamefont
  {Miao}}, \bibinfo {author} {\bibfnamefont {P.}~\bibnamefont {Charalambous}},
  \bibinfo {author} {\bibfnamefont {J.}~\bibnamefont {Kirz}}, \ and\ \bibinfo
  {author} {\bibfnamefont {D.}~\bibnamefont {Sayre}},\ }\bibfield  {title}
  {\enquote {\bibinfo {title} {Extending the methodology of x-ray
  crystallography to allow imaging of micrometre-sized non-crystalline
  specimens},}\ }\href@noop {} {\bibfield  {journal} {\bibinfo  {journal}
  {Nature}\ }\textbf {\bibinfo {volume} {400}},\ \bibinfo {pages} {342--344}
  (\bibinfo {year} {1999})}\BibitemShut {NoStop}%
\bibitem [{\citenamefont {Robinson}\ \emph {et~al.}(2001)\citenamefont
  {Robinson}, \citenamefont {Vartanyants}, \citenamefont {Williams},
  \citenamefont {Pfeifer},\ and\ \citenamefont {Pitney}}]{ref:Au_Robinson}%
  \BibitemOpen
  \bibfield  {author} {\bibinfo {author} {\bibfnamefont {I.~K.}\ \bibnamefont
  {Robinson}}, \bibinfo {author} {\bibfnamefont {I.~A.}\ \bibnamefont
  {Vartanyants}}, \bibinfo {author} {\bibfnamefont {G.}~\bibnamefont
  {Williams}}, \bibinfo {author} {\bibfnamefont {M.}~\bibnamefont {Pfeifer}}, \
  and\ \bibinfo {author} {\bibfnamefont {J.}~\bibnamefont {Pitney}},\
  }\bibfield  {title} {\enquote {\bibinfo {title} {Reconstruction of the shapes
  of gold nanocrystals using coherent x-ray diffraction},}\ }\href {\doibase
  https://doi.org/10.1103/PhysRevLett.87.195505} {\bibfield  {journal}
  {\bibinfo  {journal} {Physical review letters}\ }\textbf {\bibinfo {volume}
  {87}},\ \bibinfo {pages} {195505} (\bibinfo {year} {2001})}\BibitemShut
  {NoStop}%
\bibitem [{\citenamefont {Pfeifer}\ \emph {et~al.}(2006)\citenamefont
  {Pfeifer}, \citenamefont {Williams}, \citenamefont {Vartanyants},
  \citenamefont {Harder},\ and\ \citenamefont
  {Robinson}}]{ref:CDI_deformation}%
  \BibitemOpen
  \bibfield  {author} {\bibinfo {author} {\bibfnamefont {M.~A.}\ \bibnamefont
  {Pfeifer}}, \bibinfo {author} {\bibfnamefont {G.~J.}\ \bibnamefont
  {Williams}}, \bibinfo {author} {\bibfnamefont {I.~A.}\ \bibnamefont
  {Vartanyants}}, \bibinfo {author} {\bibfnamefont {R.}~\bibnamefont {Harder}},
  \ and\ \bibinfo {author} {\bibfnamefont {I.~K.}\ \bibnamefont {Robinson}},\
  }\bibfield  {title} {\enquote {\bibinfo {title} {Three-dimensional mapping of
  a deformation field inside a nanocrystal},}\ }\href@noop {} {\bibfield
  {journal} {\bibinfo  {journal} {Nature}\ }\textbf {\bibinfo {volume} {442}},\
  \bibinfo {pages} {63--66} (\bibinfo {year} {2006})}\BibitemShut {NoStop}%
\bibitem [{\citenamefont {Clark}\ \emph
  {et~al.}(2012{\natexlab{a}})\citenamefont {Clark}, \citenamefont {Huang},
  \citenamefont {Harder},\ and\ \citenamefont {Robinson}}]{ref:CDI_Robinson}%
  \BibitemOpen
  \bibfield  {author} {\bibinfo {author} {\bibfnamefont {J.}~\bibnamefont
  {Clark}}, \bibinfo {author} {\bibfnamefont {X.}~\bibnamefont {Huang}},
  \bibinfo {author} {\bibfnamefont {R.}~\bibnamefont {Harder}}, \ and\ \bibinfo
  {author} {\bibfnamefont {I.}~\bibnamefont {Robinson}},\ }\bibfield  {title}
  {\enquote {\bibinfo {title} {High-resolution three-dimensional partially
  coherent diffraction imaging},}\ }\href {\doibase
  https://doi.org/10.1038/ncomms1994} {\bibfield  {journal} {\bibinfo
  {journal} {Nature communications}\ }\textbf {\bibinfo {volume} {3}},\
  \bibinfo {pages} {1--6} (\bibinfo {year} {2012}{\natexlab{a}})}\BibitemShut
  {NoStop}%
\bibitem [{\citenamefont {Clark}\ \emph
  {et~al.}(2012{\natexlab{b}})\citenamefont {Clark}, \citenamefont {Huang},
  \citenamefont {Harder},\ and\ \citenamefont {Robinson}}]{ref:CDI1}%
  \BibitemOpen
  \bibfield  {author} {\bibinfo {author} {\bibfnamefont {J.}~\bibnamefont
  {Clark}}, \bibinfo {author} {\bibfnamefont {X.}~\bibnamefont {Huang}},
  \bibinfo {author} {\bibfnamefont {R.}~\bibnamefont {Harder}}, \ and\ \bibinfo
  {author} {\bibfnamefont {I.}~\bibnamefont {Robinson}},\ }\bibfield  {title}
  {\enquote {\bibinfo {title} {High-resolution three-dimensional partially
  coherent diffraction imaging},}\ }\href {\doibase 10.1038/ncomms1994}
  {\bibfield  {journal} {\bibinfo  {journal} {Nature communications}\ }\textbf
  {\bibinfo {volume} {3}},\ \bibinfo {pages} {1--6} (\bibinfo {year}
  {2012}{\natexlab{b}})}\BibitemShut {NoStop}%
\bibitem [{\citenamefont {Robinson}\ and\ \citenamefont
  {Harder}(2009)}]{ref:CDI_strain}%
  \BibitemOpen
  \bibfield  {author} {\bibinfo {author} {\bibfnamefont {I.}~\bibnamefont
  {Robinson}}\ and\ \bibinfo {author} {\bibfnamefont {R.}~\bibnamefont
  {Harder}},\ }\bibfield  {title} {\enquote {\bibinfo {title} {Coherent x-ray
  diffraction imaging of strain at the nanoscale},}\ }\href@noop {} {\bibfield
  {journal} {\bibinfo  {journal} {Nature materials}\ }\textbf {\bibinfo
  {volume} {8}},\ \bibinfo {pages} {291--298} (\bibinfo {year}
  {2009})}\BibitemShut {NoStop}%
\bibitem [{\citenamefont {Miao}, \citenamefont {Sandberg},\ and\ \citenamefont
  {Song}(2011)}]{ref:CDI_review}%
  \BibitemOpen
  \bibfield  {author} {\bibinfo {author} {\bibfnamefont {J.}~\bibnamefont
  {Miao}}, \bibinfo {author} {\bibfnamefont {R.~L.}\ \bibnamefont {Sandberg}},
  \ and\ \bibinfo {author} {\bibfnamefont {C.}~\bibnamefont {Song}},\
  }\bibfield  {title} {\enquote {\bibinfo {title} {Coherent x-ray diffraction
  imaging},}\ }\href@noop {} {\bibfield  {journal} {\bibinfo  {journal} {IEEE
  Journal of selected topics in quantum electronics}\ }\textbf {\bibinfo
  {volume} {18}},\ \bibinfo {pages} {399--410} (\bibinfo {year}
  {2011})}\BibitemShut {NoStop}%
\bibitem [{\citenamefont {Song}\ \emph {et~al.}(2008)\citenamefont {Song},
  \citenamefont {Jiang}, \citenamefont {Mancuso}, \citenamefont {Amirbekian},
  \citenamefont {Peng}, \citenamefont {Sun}, \citenamefont {Shah},
  \citenamefont {Zhou}, \citenamefont {Ishikawa},\ and\ \citenamefont
  {Miao}}]{ref:CDI_virus}%
  \BibitemOpen
  \bibfield  {author} {\bibinfo {author} {\bibfnamefont {C.}~\bibnamefont
  {Song}}, \bibinfo {author} {\bibfnamefont {H.}~\bibnamefont {Jiang}},
  \bibinfo {author} {\bibfnamefont {A.}~\bibnamefont {Mancuso}}, \bibinfo
  {author} {\bibfnamefont {B.}~\bibnamefont {Amirbekian}}, \bibinfo {author}
  {\bibfnamefont {L.}~\bibnamefont {Peng}}, \bibinfo {author} {\bibfnamefont
  {R.}~\bibnamefont {Sun}}, \bibinfo {author} {\bibfnamefont {S.~S.}\
  \bibnamefont {Shah}}, \bibinfo {author} {\bibfnamefont {Z.~H.}\ \bibnamefont
  {Zhou}}, \bibinfo {author} {\bibfnamefont {T.}~\bibnamefont {Ishikawa}}, \
  and\ \bibinfo {author} {\bibfnamefont {J.}~\bibnamefont {Miao}},\ }\bibfield
  {title} {\enquote {\bibinfo {title} {Quantitative imaging of single,
  unstained viruses with coherent x rays},}\ }\href@noop {} {\bibfield
  {journal} {\bibinfo  {journal} {Physical review letters}\ }\textbf {\bibinfo
  {volume} {101}},\ \bibinfo {pages} {158101} (\bibinfo {year}
  {2008})}\BibitemShut {NoStop}%
\bibitem [{\citenamefont {Miao}\ \emph {et~al.}(2003)\citenamefont {Miao},
  \citenamefont {Hodgson}, \citenamefont {Ishikawa}, \citenamefont {Larabell},
  \citenamefont {LeGros},\ and\ \citenamefont {Nishino}}]{ref:CDI_bacteria}%
  \BibitemOpen
  \bibfield  {author} {\bibinfo {author} {\bibfnamefont {J.}~\bibnamefont
  {Miao}}, \bibinfo {author} {\bibfnamefont {K.~O.}\ \bibnamefont {Hodgson}},
  \bibinfo {author} {\bibfnamefont {T.}~\bibnamefont {Ishikawa}}, \bibinfo
  {author} {\bibfnamefont {C.~A.}\ \bibnamefont {Larabell}}, \bibinfo {author}
  {\bibfnamefont {M.~A.}\ \bibnamefont {LeGros}}, \ and\ \bibinfo {author}
  {\bibfnamefont {Y.}~\bibnamefont {Nishino}},\ }\bibfield  {title} {\enquote
  {\bibinfo {title} {Imaging whole escherichia coli bacteria by using
  single-particle x-ray diffraction},}\ }\href@noop {} {\bibfield  {journal}
  {\bibinfo  {journal} {Proceedings of the National Academy of Sciences}\
  }\textbf {\bibinfo {volume} {100}},\ \bibinfo {pages} {110--112} (\bibinfo
  {year} {2003})}\BibitemShut {NoStop}%
\bibitem [{\citenamefont {Miao}\ \emph {et~al.}(2005)\citenamefont {Miao},
  \citenamefont {Nishino}, \citenamefont {Kohmura}, \citenamefont {Johnson},
  \citenamefont {Song}, \citenamefont {Risbud},\ and\ \citenamefont
  {Ishikawa}}]{ref:CDI_quantum}%
  \BibitemOpen
  \bibfield  {author} {\bibinfo {author} {\bibfnamefont {J.}~\bibnamefont
  {Miao}}, \bibinfo {author} {\bibfnamefont {Y.}~\bibnamefont {Nishino}},
  \bibinfo {author} {\bibfnamefont {Y.}~\bibnamefont {Kohmura}}, \bibinfo
  {author} {\bibfnamefont {B.}~\bibnamefont {Johnson}}, \bibinfo {author}
  {\bibfnamefont {C.}~\bibnamefont {Song}}, \bibinfo {author} {\bibfnamefont
  {S.~H.}\ \bibnamefont {Risbud}}, \ and\ \bibinfo {author} {\bibfnamefont
  {T.}~\bibnamefont {Ishikawa}},\ }\bibfield  {title} {\enquote {\bibinfo
  {title} {Quantitative image reconstruction of gan quantum dots from
  oversampled diffraction intensities alone},}\ }\href@noop {} {\bibfield
  {journal} {\bibinfo  {journal} {Physical review letters}\ }\textbf {\bibinfo
  {volume} {95}},\ \bibinfo {pages} {085503} (\bibinfo {year}
  {2005})}\BibitemShut {NoStop}%
\bibitem [{\citenamefont {Williams}\ \emph {et~al.}(2008)\citenamefont
  {Williams}, \citenamefont {Hanssen}, \citenamefont {Peele}, \citenamefont
  {Pfeifer}, \citenamefont {Clark}, \citenamefont {Abbey}, \citenamefont
  {Cadenazzi}, \citenamefont {de~Jonge}, \citenamefont {Vogt}, \citenamefont
  {Tilley} \emph {et~al.}}]{ref:CDI_blood}%
  \BibitemOpen
  \bibfield  {author} {\bibinfo {author} {\bibfnamefont {G.~J.}\ \bibnamefont
  {Williams}}, \bibinfo {author} {\bibfnamefont {E.}~\bibnamefont {Hanssen}},
  \bibinfo {author} {\bibfnamefont {A.~G.}\ \bibnamefont {Peele}}, \bibinfo
  {author} {\bibfnamefont {M.~A.}\ \bibnamefont {Pfeifer}}, \bibinfo {author}
  {\bibfnamefont {J.}~\bibnamefont {Clark}}, \bibinfo {author} {\bibfnamefont
  {B.}~\bibnamefont {Abbey}}, \bibinfo {author} {\bibfnamefont
  {G.}~\bibnamefont {Cadenazzi}}, \bibinfo {author} {\bibfnamefont {M.~D.}\
  \bibnamefont {de~Jonge}}, \bibinfo {author} {\bibfnamefont {S.}~\bibnamefont
  {Vogt}}, \bibinfo {author} {\bibfnamefont {L.}~\bibnamefont {Tilley}},  \emph
  {et~al.},\ }\bibfield  {title} {\enquote {\bibinfo {title} {High-resolution
  x-ray imaging of plasmodium falciparum-infected red blood cells},}\
  }\href@noop {} {\bibfield  {journal} {\bibinfo  {journal} {Cytometry Part A:
  The Journal of the International Society for Analytical Cytology}\ }\textbf
  {\bibinfo {volume} {73}},\ \bibinfo {pages} {949--957} (\bibinfo {year}
  {2008})}\BibitemShut {NoStop}%
\bibitem [{\citenamefont {Nishino}\ \emph {et~al.}(2009)\citenamefont
  {Nishino}, \citenamefont {Takahashi}, \citenamefont {Imamoto}, \citenamefont
  {Ishikawa},\ and\ \citenamefont {Maeshima}}]{ref:CDI_chromosome}%
  \BibitemOpen
  \bibfield  {author} {\bibinfo {author} {\bibfnamefont {Y.}~\bibnamefont
  {Nishino}}, \bibinfo {author} {\bibfnamefont {Y.}~\bibnamefont {Takahashi}},
  \bibinfo {author} {\bibfnamefont {N.}~\bibnamefont {Imamoto}}, \bibinfo
  {author} {\bibfnamefont {T.}~\bibnamefont {Ishikawa}}, \ and\ \bibinfo
  {author} {\bibfnamefont {K.}~\bibnamefont {Maeshima}},\ }\bibfield  {title}
  {\enquote {\bibinfo {title} {Three-dimensional visualization of a human
  chromosome using coherent x-ray diffraction},}\ }\href@noop {} {\bibfield
  {journal} {\bibinfo  {journal} {Physical review letters}\ }\textbf {\bibinfo
  {volume} {102}},\ \bibinfo {pages} {018101} (\bibinfo {year}
  {2009})}\BibitemShut {NoStop}%
\bibitem [{\citenamefont {Huang}\ \emph {et~al.}(2012)\citenamefont {Huang},
  \citenamefont {Harder}, \citenamefont {Leake}, \citenamefont {Clark},\ and\
  \citenamefont {Robinson}}]{ref:CDI_ZnO}%
  \BibitemOpen
  \bibfield  {author} {\bibinfo {author} {\bibfnamefont {X.}~\bibnamefont
  {Huang}}, \bibinfo {author} {\bibfnamefont {R.}~\bibnamefont {Harder}},
  \bibinfo {author} {\bibfnamefont {S.}~\bibnamefont {Leake}}, \bibinfo
  {author} {\bibfnamefont {J.}~\bibnamefont {Clark}}, \ and\ \bibinfo {author}
  {\bibfnamefont {I.}~\bibnamefont {Robinson}},\ }\bibfield  {title} {\enquote
  {\bibinfo {title} {Three-dimensional bragg coherent diffraction imaging of an
  extended zno crystal},}\ }\href@noop {} {\bibfield  {journal} {\bibinfo
  {journal} {Journal of applied crystallography}\ }\textbf {\bibinfo {volume}
  {45}},\ \bibinfo {pages} {778--784} (\bibinfo {year} {2012})}\BibitemShut
  {NoStop}%
\bibitem [{\citenamefont {Clark}\ \emph
  {et~al.}(2012{\natexlab{c}})\citenamefont {Clark}, \citenamefont {Huang},
  \citenamefont {Harder},\ and\ \citenamefont {Robinson}}]{ref:CDI_PC}%
  \BibitemOpen
  \bibfield  {author} {\bibinfo {author} {\bibfnamefont {J.}~\bibnamefont
  {Clark}}, \bibinfo {author} {\bibfnamefont {X.}~\bibnamefont {Huang}},
  \bibinfo {author} {\bibfnamefont {R.}~\bibnamefont {Harder}}, \ and\ \bibinfo
  {author} {\bibfnamefont {I.}~\bibnamefont {Robinson}},\ }\bibfield  {title}
  {\enquote {\bibinfo {title} {High-resolution three-dimensional partially
  coherent diffraction imaging},}\ }\href@noop {} {\bibfield  {journal}
  {\bibinfo  {journal} {Nature communications}\ }\textbf {\bibinfo {volume}
  {3}},\ \bibinfo {pages} {1--6} (\bibinfo {year}
  {2012}{\natexlab{c}})}\BibitemShut {NoStop}%
\bibitem [{\citenamefont {Hofmann}\ \emph {et~al.}(2017)\citenamefont
  {Hofmann}, \citenamefont {Tarleton}, \citenamefont {Harder}, \citenamefont
  {Phillips}, \citenamefont {Ma}, \citenamefont {Clark}, \citenamefont
  {Robinson}, \citenamefont {Abbey}, \citenamefont {Liu},\ and\ \citenamefont
  {Beck}}]{ref:BCDI:Felix1}%
  \BibitemOpen
  \bibfield  {author} {\bibinfo {author} {\bibfnamefont {F.}~\bibnamefont
  {Hofmann}}, \bibinfo {author} {\bibfnamefont {E.}~\bibnamefont {Tarleton}},
  \bibinfo {author} {\bibfnamefont {R.~J.}\ \bibnamefont {Harder}}, \bibinfo
  {author} {\bibfnamefont {N.~W.}\ \bibnamefont {Phillips}}, \bibinfo {author}
  {\bibfnamefont {P.-W.}\ \bibnamefont {Ma}}, \bibinfo {author} {\bibfnamefont
  {J.~N.}\ \bibnamefont {Clark}}, \bibinfo {author} {\bibfnamefont {I.~K.}\
  \bibnamefont {Robinson}}, \bibinfo {author} {\bibfnamefont {B.}~\bibnamefont
  {Abbey}}, \bibinfo {author} {\bibfnamefont {W.}~\bibnamefont {Liu}}, \ and\
  \bibinfo {author} {\bibfnamefont {C.~E.}\ \bibnamefont {Beck}},\ }\bibfield
  {title} {\enquote {\bibinfo {title} {3d lattice distortions and defect
  structures in ion-implanted nano-crystals},}\ }\href@noop {} {\bibfield
  {journal} {\bibinfo  {journal} {Scientific reports}\ }\textbf {\bibinfo
  {volume} {7}},\ \bibinfo {pages} {45993} (\bibinfo {year}
  {2017})}\BibitemShut {NoStop}%
\bibitem [{\citenamefont {Cherukara}\ \emph {et~al.}(2018)\citenamefont
  {Cherukara}, \citenamefont {Pokharel}, \citenamefont {O’Leary},
  \citenamefont {Baldwin}, \citenamefont {Maxey}, \citenamefont {Cha},
  \citenamefont {Maser}, \citenamefont {Harder}, \citenamefont {Fensin},\ and\
  \citenamefont {Sandberg}}]{ref:CDI_RP}%
  \BibitemOpen
  \bibfield  {author} {\bibinfo {author} {\bibfnamefont {M.~J.}\ \bibnamefont
  {Cherukara}}, \bibinfo {author} {\bibfnamefont {R.}~\bibnamefont {Pokharel}},
  \bibinfo {author} {\bibfnamefont {T.~S.}\ \bibnamefont {O’Leary}}, \bibinfo
  {author} {\bibfnamefont {J.~K.}\ \bibnamefont {Baldwin}}, \bibinfo {author}
  {\bibfnamefont {E.}~\bibnamefont {Maxey}}, \bibinfo {author} {\bibfnamefont
  {W.}~\bibnamefont {Cha}}, \bibinfo {author} {\bibfnamefont {J.}~\bibnamefont
  {Maser}}, \bibinfo {author} {\bibfnamefont {R.~J.}\ \bibnamefont {Harder}},
  \bibinfo {author} {\bibfnamefont {S.~J.}\ \bibnamefont {Fensin}}, \ and\
  \bibinfo {author} {\bibfnamefont {R.~L.}\ \bibnamefont {Sandberg}},\
  }\bibfield  {title} {\enquote {\bibinfo {title} {Three-dimensional x-ray
  diffraction imaging of dislocations in polycrystalline metals under tensile
  loading},}\ }\href {\doibase 10.1038/s41467-018-06166-5} {\bibfield
  {journal} {\bibinfo  {journal} {Nature communications}\ }\textbf {\bibinfo
  {volume} {9}},\ \bibinfo {pages} {1--6} (\bibinfo {year} {2018})}\BibitemShut
  {NoStop}%
\bibitem [{\citenamefont {Miao}, \citenamefont {Sayre},\ and\ \citenamefont
  {Chapman}(1998)}]{ref:CDI_phase_1}%
  \BibitemOpen
  \bibfield  {author} {\bibinfo {author} {\bibfnamefont {J.}~\bibnamefont
  {Miao}}, \bibinfo {author} {\bibfnamefont {D.}~\bibnamefont {Sayre}}, \ and\
  \bibinfo {author} {\bibfnamefont {H.}~\bibnamefont {Chapman}},\ }\bibfield
  {title} {\enquote {\bibinfo {title} {Phase retrieval from the magnitude of
  the fourier transforms of nonperiodic objects},}\ }\href@noop {} {\bibfield
  {journal} {\bibinfo  {journal} {JOSA A}\ }\textbf {\bibinfo {volume} {15}},\
  \bibinfo {pages} {1662--1669} (\bibinfo {year} {1998})}\BibitemShut {NoStop}%
\bibitem [{\citenamefont {Fienup}(1978)}]{ref:CDI_phase_2}%
  \BibitemOpen
  \bibfield  {author} {\bibinfo {author} {\bibfnamefont {J.~R.}\ \bibnamefont
  {Fienup}},\ }\bibfield  {title} {\enquote {\bibinfo {title} {Reconstruction
  of an object from the modulus of its fourier transform},}\ }\href@noop {}
  {\bibfield  {journal} {\bibinfo  {journal} {Optics letters}\ }\textbf
  {\bibinfo {volume} {3}},\ \bibinfo {pages} {27--29} (\bibinfo {year}
  {1978})}\BibitemShut {NoStop}%
\bibitem [{\citenamefont {Fienup}(1982)}]{ref:CDI_phase_3}%
  \BibitemOpen
  \bibfield  {author} {\bibinfo {author} {\bibfnamefont {J.~R.}\ \bibnamefont
  {Fienup}},\ }\bibfield  {title} {\enquote {\bibinfo {title} {Phase retrieval
  algorithms: a comparison},}\ }\href@noop {} {\bibfield  {journal} {\bibinfo
  {journal} {Applied optics}\ }\textbf {\bibinfo {volume} {21}},\ \bibinfo
  {pages} {2758--2769} (\bibinfo {year} {1982})}\BibitemShut {NoStop}%
\bibitem [{\citenamefont {Elser}(2003)}]{ref:CDI_phase_4}%
  \BibitemOpen
  \bibfield  {author} {\bibinfo {author} {\bibfnamefont {V.}~\bibnamefont
  {Elser}},\ }\bibfield  {title} {\enquote {\bibinfo {title} {Solution of the
  crystallographic phase problem by iterated projections},}\ }\href@noop {}
  {\bibfield  {journal} {\bibinfo  {journal} {Acta Crystallographica Section A:
  Foundations of Crystallography}\ }\textbf {\bibinfo {volume} {59}},\ \bibinfo
  {pages} {201--209} (\bibinfo {year} {2003})}\BibitemShut {NoStop}%
\bibitem [{\citenamefont {Chen}\ \emph {et~al.}(2007)\citenamefont {Chen},
  \citenamefont {Miao}, \citenamefont {Wang},\ and\ \citenamefont
  {Lee}}]{ref:CDI_phase_5}%
  \BibitemOpen
  \bibfield  {author} {\bibinfo {author} {\bibfnamefont {C.-C.}\ \bibnamefont
  {Chen}}, \bibinfo {author} {\bibfnamefont {J.}~\bibnamefont {Miao}}, \bibinfo
  {author} {\bibfnamefont {C.}~\bibnamefont {Wang}}, \ and\ \bibinfo {author}
  {\bibfnamefont {T.}~\bibnamefont {Lee}},\ }\bibfield  {title} {\enquote
  {\bibinfo {title} {Application of optimization technique to noncrystalline
  x-ray diffraction microscopy: Guided hybrid input-output method},}\
  }\href@noop {} {\bibfield  {journal} {\bibinfo  {journal} {Physical Review
  B}\ }\textbf {\bibinfo {volume} {76}},\ \bibinfo {pages} {064113} (\bibinfo
  {year} {2007})}\BibitemShut {NoStop}%
\bibitem [{\citenamefont {Marchesini}(2007)}]{ref:CDI_phase_6}%
  \BibitemOpen
  \bibfield  {author} {\bibinfo {author} {\bibfnamefont {S.}~\bibnamefont
  {Marchesini}},\ }\bibfield  {title} {\enquote {\bibinfo {title} {Invited
  article: A unified evaluation of iterative projection algorithms for phase
  retrieval},}\ }\href@noop {} {\bibfield  {journal} {\bibinfo  {journal}
  {Review of scientific instruments}\ }\textbf {\bibinfo {volume} {78}},\
  \bibinfo {pages} {011301} (\bibinfo {year} {2007})}\BibitemShut {NoStop}%
\bibitem [{\citenamefont {Bauschke}, \citenamefont {Combettes},\ and\
  \citenamefont {Luke}(2003)}]{ref:CDI_phase_7}%
  \BibitemOpen
  \bibfield  {author} {\bibinfo {author} {\bibfnamefont {H.~H.}\ \bibnamefont
  {Bauschke}}, \bibinfo {author} {\bibfnamefont {P.~L.}\ \bibnamefont
  {Combettes}}, \ and\ \bibinfo {author} {\bibfnamefont {D.~R.}\ \bibnamefont
  {Luke}},\ }\bibfield  {title} {\enquote {\bibinfo {title} {Hybrid
  projection--reflection method for phase retrieval},}\ }\href@noop {}
  {\bibfield  {journal} {\bibinfo  {journal} {JOSA A}\ }\textbf {\bibinfo
  {volume} {20}},\ \bibinfo {pages} {1025--1034} (\bibinfo {year}
  {2003})}\BibitemShut {NoStop}%
\bibitem [{\citenamefont {Luke}(2004)}]{ref:CDI_phase_8}%
  \BibitemOpen
  \bibfield  {author} {\bibinfo {author} {\bibfnamefont {D.~R.}\ \bibnamefont
  {Luke}},\ }\bibfield  {title} {\enquote {\bibinfo {title} {Relaxed averaged
  alternating reflections for diffraction imaging},}\ }\href@noop {} {\bibfield
   {journal} {\bibinfo  {journal} {Inverse problems}\ }\textbf {\bibinfo
  {volume} {21}},\ \bibinfo {pages} {37} (\bibinfo {year} {2004})}\BibitemShut
  {NoStop}%
\bibitem [{\citenamefont {Sayre}(1980)}]{ref:CDI_phase_9}%
  \BibitemOpen
  \bibfield  {author} {\bibinfo {author} {\bibfnamefont {D.}~\bibnamefont
  {Sayre}},\ }\bibfield  {title} {\enquote {\bibinfo {title} {Imaging processes
  and coherence in physics},}\ }\href@noop {} {\bibfield  {journal} {\bibinfo
  {journal} {Springer Lecture Notes in Physics}\ }\textbf {\bibinfo {volume}
  {112}},\ \bibinfo {pages} {229--235} (\bibinfo {year} {1980})}\BibitemShut
  {NoStop}%
\bibitem [{\citenamefont {Colombo}\ \emph {et~al.}(2017)\citenamefont
  {Colombo}, \citenamefont {Galli}, \citenamefont {De~Caro}, \citenamefont
  {Scattarella},\ and\ \citenamefont {Carlino}}]{ref:CDI_mematic}%
  \BibitemOpen
  \bibfield  {author} {\bibinfo {author} {\bibfnamefont {A.}~\bibnamefont
  {Colombo}}, \bibinfo {author} {\bibfnamefont {D.~E.}\ \bibnamefont {Galli}},
  \bibinfo {author} {\bibfnamefont {L.}~\bibnamefont {De~Caro}}, \bibinfo
  {author} {\bibfnamefont {F.}~\bibnamefont {Scattarella}}, \ and\ \bibinfo
  {author} {\bibfnamefont {E.}~\bibnamefont {Carlino}},\ }\bibfield  {title}
  {\enquote {\bibinfo {title} {Facing the phase problem in coherent diffractive
  imaging via memetic algorithms},}\ }\href@noop {} {\bibfield  {journal}
  {\bibinfo  {journal} {Scientific reports}\ }\textbf {\bibinfo {volume} {7}},\
  \bibinfo {pages} {1--12} (\bibinfo {year} {2017})}\BibitemShut {NoStop}%
\bibitem [{\citenamefont {Fevola}\ \emph {et~al.}(2020)\citenamefont {Fevola},
  \citenamefont {Bergb{\"a}ck~Knudsen}, \citenamefont {Ramos}, \citenamefont
  {Carbone},\ and\ \citenamefont {Wenzel~Andreasen}}]{ref:CDI_raytrace}%
  \BibitemOpen
  \bibfield  {author} {\bibinfo {author} {\bibfnamefont {G.}~\bibnamefont
  {Fevola}}, \bibinfo {author} {\bibfnamefont {E.}~\bibnamefont
  {Bergb{\"a}ck~Knudsen}}, \bibinfo {author} {\bibfnamefont {T.}~\bibnamefont
  {Ramos}}, \bibinfo {author} {\bibfnamefont {D.}~\bibnamefont {Carbone}}, \
  and\ \bibinfo {author} {\bibfnamefont {J.}~\bibnamefont {Wenzel~Andreasen}},\
  }\bibfield  {title} {\enquote {\bibinfo {title} {A monte carlo ray-tracing
  simulation of coherent x-ray diffractive imaging},}\ }\href@noop {}
  {\bibfield  {journal} {\bibinfo  {journal} {Journal of Synchrotron
  Radiation}\ }\textbf {\bibinfo {volume} {27}} (\bibinfo {year}
  {2020})}\BibitemShut {NoStop}%
\bibitem [{\citenamefont {Maddali}\ \emph {et~al.}(2020)\citenamefont
  {Maddali}, \citenamefont {Li}, \citenamefont {Pateras}, \citenamefont
  {Timbie}, \citenamefont {Delegan}, \citenamefont {Crook}, \citenamefont
  {Lee}, \citenamefont {Calvo-Almazan}, \citenamefont {Sheyfer}, \citenamefont
  {Cha} \emph {et~al.}}]{ref:BCDI:Sid2}%
  \BibitemOpen
  \bibfield  {author} {\bibinfo {author} {\bibfnamefont {S.}~\bibnamefont
  {Maddali}}, \bibinfo {author} {\bibfnamefont {P.}~\bibnamefont {Li}},
  \bibinfo {author} {\bibfnamefont {A.}~\bibnamefont {Pateras}}, \bibinfo
  {author} {\bibfnamefont {D.}~\bibnamefont {Timbie}}, \bibinfo {author}
  {\bibfnamefont {N.}~\bibnamefont {Delegan}}, \bibinfo {author} {\bibfnamefont
  {A.}~\bibnamefont {Crook}}, \bibinfo {author} {\bibfnamefont
  {H.}~\bibnamefont {Lee}}, \bibinfo {author} {\bibfnamefont {I.}~\bibnamefont
  {Calvo-Almazan}}, \bibinfo {author} {\bibfnamefont {D.}~\bibnamefont
  {Sheyfer}}, \bibinfo {author} {\bibfnamefont {W.}~\bibnamefont {Cha}},  \emph
  {et~al.},\ }\bibfield  {title} {\enquote {\bibinfo {title} {General
  approaches for shear-correcting coordinate transformations in bragg coherent
  diffraction imaging. part i},}\ }\href@noop {} {\bibfield  {journal}
  {\bibinfo  {journal} {Journal of Applied Crystallography}\ }\textbf {\bibinfo
  {volume} {53}} (\bibinfo {year} {2020})}\BibitemShut {NoStop}%
\bibitem [{\citenamefont {Li}\ \emph {et~al.}(2020)\citenamefont {Li},
  \citenamefont {Maddali}, \citenamefont {Pateras}, \citenamefont
  {Calvo-Almazan}, \citenamefont {Hruszkewycz}, \citenamefont {Cha},
  \citenamefont {Chamard},\ and\ \citenamefont {Allain}}]{ref:BCDI:Sid3}%
  \BibitemOpen
  \bibfield  {author} {\bibinfo {author} {\bibfnamefont {P.}~\bibnamefont
  {Li}}, \bibinfo {author} {\bibfnamefont {S.}~\bibnamefont {Maddali}},
  \bibinfo {author} {\bibfnamefont {A.}~\bibnamefont {Pateras}}, \bibinfo
  {author} {\bibfnamefont {I.}~\bibnamefont {Calvo-Almazan}}, \bibinfo {author}
  {\bibfnamefont {S.~O.}\ \bibnamefont {Hruszkewycz}}, \bibinfo {author}
  {\bibfnamefont {W.}~\bibnamefont {Cha}}, \bibinfo {author} {\bibfnamefont
  {V.}~\bibnamefont {Chamard}}, \ and\ \bibinfo {author} {\bibfnamefont
  {M.}~\bibnamefont {Allain}},\ }\bibfield  {title} {\enquote {\bibinfo {title}
  {General approaches for shear-correcting coordinate transformations in bragg
  coherent diffraction imaging. part ii},}\ }\href@noop {} {\bibfield
  {journal} {\bibinfo  {journal} {Journal of Applied Crystallography}\ }\textbf
  {\bibinfo {volume} {53}} (\bibinfo {year} {2020})}\BibitemShut {NoStop}%
\bibitem [{\citenamefont {Godard}\ \emph {et~al.}(2012)\citenamefont {Godard},
  \citenamefont {Allain}, \citenamefont {Chamard},\ and\ \citenamefont
  {Rodenburg}}]{ref:CDInoise}%
  \BibitemOpen
  \bibfield  {author} {\bibinfo {author} {\bibfnamefont {P.}~\bibnamefont
  {Godard}}, \bibinfo {author} {\bibfnamefont {M.}~\bibnamefont {Allain}},
  \bibinfo {author} {\bibfnamefont {V.}~\bibnamefont {Chamard}}, \ and\
  \bibinfo {author} {\bibfnamefont {J.}~\bibnamefont {Rodenburg}},\ }\bibfield
  {title} {\enquote {\bibinfo {title} {Noise models for low counting rate
  coherent diffraction imaging},}\ }\href@noop {} {\bibfield  {journal}
  {\bibinfo  {journal} {Optics express}\ }\textbf {\bibinfo {volume} {20}},\
  \bibinfo {pages} {25914--25934} (\bibinfo {year} {2012})}\BibitemShut
  {NoStop}%
\bibitem [{\citenamefont {Li}\ \emph {et~al.}(2019)\citenamefont {Li},
  \citenamefont {Dee}, \citenamefont {Khatami},\ and\ \citenamefont
  {Johnston}}]{ref:ML_montecarlo}%
  \BibitemOpen
  \bibfield  {author} {\bibinfo {author} {\bibfnamefont {S.}~\bibnamefont
  {Li}}, \bibinfo {author} {\bibfnamefont {P.~M.}\ \bibnamefont {Dee}},
  \bibinfo {author} {\bibfnamefont {E.}~\bibnamefont {Khatami}}, \ and\
  \bibinfo {author} {\bibfnamefont {S.}~\bibnamefont {Johnston}},\ }\bibfield
  {title} {\enquote {\bibinfo {title} {Accelerating lattice quantum monte carlo
  simulations using artificial neural networks: Application to the holstein
  model},}\ }\href@noop {} {\bibfield  {journal} {\bibinfo  {journal} {Physical
  Review B}\ }\textbf {\bibinfo {volume} {100}},\ \bibinfo {pages} {020302}
  (\bibinfo {year} {2019})}\BibitemShut {NoStop}%
\bibitem [{\citenamefont {Rrapaj}\ and\ \citenamefont
  {Roggero}(2020)}]{ref:ML_manybody}%
  \BibitemOpen
  \bibfield  {author} {\bibinfo {author} {\bibfnamefont {E.}~\bibnamefont
  {Rrapaj}}\ and\ \bibinfo {author} {\bibfnamefont {A.}~\bibnamefont
  {Roggero}},\ }\bibfield  {title} {\enquote {\bibinfo {title} {Exact
  representations of many body interactions with rbm neural networks},}\
  }\href@noop {} {\bibfield  {journal} {\bibinfo  {journal} {arXiv preprint
  arXiv:2005.03568}\ } (\bibinfo {year} {2020})}\BibitemShut {NoStop}%
\bibitem [{\citenamefont {Shimobaba}, \citenamefont {Kakue},\ and\
  \citenamefont {Ito}(2018)}]{ref:ML_hologram}%
  \BibitemOpen
  \bibfield  {author} {\bibinfo {author} {\bibfnamefont {T.}~\bibnamefont
  {Shimobaba}}, \bibinfo {author} {\bibfnamefont {T.}~\bibnamefont {Kakue}}, \
  and\ \bibinfo {author} {\bibfnamefont {T.}~\bibnamefont {Ito}},\ }\bibfield
  {title} {\enquote {\bibinfo {title} {Convolutional neural network-based
  regression for depth prediction in digital holography},}\ }in\ \href@noop {}
  {\emph {\bibinfo {booktitle} {2018 IEEE 27th International Symposium on
  Industrial Electronics (ISIE)}}}\ (\bibinfo {organization} {IEEE},\ \bibinfo
  {year} {2018})\ pp.\ \bibinfo {pages} {1323--1326}\BibitemShut {NoStop}%
\bibitem [{\citenamefont {Scheinker}\ \emph {et~al.}(2018)\citenamefont
  {Scheinker}, \citenamefont {Edelen}, \citenamefont {Bohler}, \citenamefont
  {Emma},\ and\ \citenamefont {Lutman}}]{ref:ESML_AS}%
  \BibitemOpen
  \bibfield  {author} {\bibinfo {author} {\bibfnamefont {A.}~\bibnamefont
  {Scheinker}}, \bibinfo {author} {\bibfnamefont {A.}~\bibnamefont {Edelen}},
  \bibinfo {author} {\bibfnamefont {D.}~\bibnamefont {Bohler}}, \bibinfo
  {author} {\bibfnamefont {C.}~\bibnamefont {Emma}}, \ and\ \bibinfo {author}
  {\bibfnamefont {A.}~\bibnamefont {Lutman}},\ }\bibfield  {title} {\enquote
  {\bibinfo {title} {Demonstration of model-independent control of the
  longitudinal phase space of electron beams in the linac-coherent light source
  with femtosecond resolution},}\ }\href {\doibase
  https://doi.org/10.1103/PhysRevLett.121.044801} {\bibfield  {journal}
  {\bibinfo  {journal} {Physical review letters}\ }\textbf {\bibinfo {volume}
  {121}},\ \bibinfo {pages} {044801} (\bibinfo {year} {2018})}\BibitemShut
  {NoStop}%
\bibitem [{\citenamefont {Cherukara}, \citenamefont {Nashed},\ and\
  \citenamefont {Harder}(2018)}]{ref:realtimeCDI}%
  \BibitemOpen
  \bibfield  {author} {\bibinfo {author} {\bibfnamefont {M.~J.}\ \bibnamefont
  {Cherukara}}, \bibinfo {author} {\bibfnamefont {Y.~S.}\ \bibnamefont
  {Nashed}}, \ and\ \bibinfo {author} {\bibfnamefont {R.~J.}\ \bibnamefont
  {Harder}},\ }\bibfield  {title} {\enquote {\bibinfo {title} {Real-time
  coherent diffraction inversion using deep generative networks},}\ }\href@noop
  {} {\bibfield  {journal} {\bibinfo  {journal} {Scientific reports}\ }\textbf
  {\bibinfo {volume} {8}},\ \bibinfo {pages} {1--8} (\bibinfo {year}
  {2018})}\BibitemShut {NoStop}%
\bibitem [{\citenamefont {Shen}\ \emph {et~al.}(2019)\citenamefont {Shen},
  \citenamefont {Pokharel}, \citenamefont {Nizolek}, \citenamefont {Kumar},\
  and\ \citenamefont {Lookman}}]{ref:EBSD_RP}%
  \BibitemOpen
  \bibfield  {author} {\bibinfo {author} {\bibfnamefont {Y.-F.}\ \bibnamefont
  {Shen}}, \bibinfo {author} {\bibfnamefont {R.}~\bibnamefont {Pokharel}},
  \bibinfo {author} {\bibfnamefont {T.~J.}\ \bibnamefont {Nizolek}}, \bibinfo
  {author} {\bibfnamefont {A.}~\bibnamefont {Kumar}}, \ and\ \bibinfo {author}
  {\bibfnamefont {T.}~\bibnamefont {Lookman}},\ }\bibfield  {title} {\enquote
  {\bibinfo {title} {Convolutional neural network-based method for real-time
  orientation indexing of measured electron backscatter diffraction
  patterns},}\ }\href {\doibase https://doi.org/10.1016/j.actamat.2019.03.026}
  {\bibfield  {journal} {\bibinfo  {journal} {Acta Materialia}\ }\textbf
  {\bibinfo {volume} {170}},\ \bibinfo {pages} {118--131} (\bibinfo {year}
  {2019})}\BibitemShut {NoStop}%
\bibitem [{\citenamefont {Kandel}\ \emph {et~al.}(2019)\citenamefont {Kandel},
  \citenamefont {Maddali}, \citenamefont {Allain}, \citenamefont {Hruszkewycz},
  \citenamefont {Jacobsen},\ and\ \citenamefont {Nashed}}]{ref:autoDiff}%
  \BibitemOpen
  \bibfield  {author} {\bibinfo {author} {\bibfnamefont {S.}~\bibnamefont
  {Kandel}}, \bibinfo {author} {\bibfnamefont {S.}~\bibnamefont {Maddali}},
  \bibinfo {author} {\bibfnamefont {M.}~\bibnamefont {Allain}}, \bibinfo
  {author} {\bibfnamefont {S.~O.}\ \bibnamefont {Hruszkewycz}}, \bibinfo
  {author} {\bibfnamefont {C.}~\bibnamefont {Jacobsen}}, \ and\ \bibinfo
  {author} {\bibfnamefont {Y.~S.}\ \bibnamefont {Nashed}},\ }\bibfield  {title}
  {\enquote {\bibinfo {title} {Using automatic differentiation as a general
  framework for ptychographic reconstruction},}\ }\href@noop {} {\bibfield
  {journal} {\bibinfo  {journal} {Optics express}\ }\textbf {\bibinfo {volume}
  {27}},\ \bibinfo {pages} {18653--18672} (\bibinfo {year} {2019})}\BibitemShut
  {NoStop}%
\bibitem [{\citenamefont {Scheinker}(2013)}]{ref:ES_Bounded}%
  \BibitemOpen
  \bibfield  {author} {\bibinfo {author} {\bibfnamefont {A.}~\bibnamefont
  {Scheinker}},\ }\bibfield  {title} {\enquote {\bibinfo {title} {Model
  independent beam tuning},}\ }in\ \href
  {https://accelconf.web.cern.ch/IPAC2013/papers/tupwa068.pdf?n=IPAC2013/papers/tupwa068.pdf}
  {\emph {\bibinfo {booktitle} {Proceedings of the 2013 International Particle
  Accelerator Conference, Shanghai, China}}}\ (\bibinfo {year} {2013})\ p.\
  \bibinfo {pages} {TUPWA068}\BibitemShut {NoStop}%
\bibitem [{\citenamefont {Scheinker}\ and\ \citenamefont
  {Scheinker}(2016)}]{ref:ES_nonC2}%
  \BibitemOpen
  \bibfield  {author} {\bibinfo {author} {\bibfnamefont {A.}~\bibnamefont
  {Scheinker}}\ and\ \bibinfo {author} {\bibfnamefont {D.}~\bibnamefont
  {Scheinker}},\ }\bibfield  {title} {\enquote {\bibinfo {title} {Bounded
  extremum seeking with discontinuous dithers},}\ }\href {\doibase
  https://doi.org/10.1016/j.automatica.2016.02.023} {\bibfield  {journal}
  {\bibinfo  {journal} {Automatica}\ }\textbf {\bibinfo {volume} {69}},\
  \bibinfo {pages} {250--257} (\bibinfo {year} {2016})}\BibitemShut {NoStop}%
\bibitem [{\citenamefont {Scheinker}\ and\ \citenamefont
  {Scheinker}(2020)}]{ref:ES_opt}%
  \BibitemOpen
  \bibfield  {author} {\bibinfo {author} {\bibfnamefont {A.}~\bibnamefont
  {Scheinker}}\ and\ \bibinfo {author} {\bibfnamefont {D.}~\bibnamefont
  {Scheinker}},\ }\bibfield  {title} {\enquote {\bibinfo {title} {Extremum
  seeking for optimal control problems with unknown time-varying systems and
  unknown objective functions},}\ }\href {\doibase
  https://doi.org/10.1002/acs.3097} {\bibfield  {journal} {\bibinfo  {journal}
  {International Journal of Adaptive Control and Signal Processing}\ }
  (\bibinfo {year} {2020}),\ https://doi.org/10.1002/acs.3097}\BibitemShut
  {NoStop}%
\bibitem [{\citenamefont {Pokharel}\ \emph {et~al.}(2015)\citenamefont
  {Pokharel}, \citenamefont {Lind}, \citenamefont {Li}, \citenamefont
  {Kenesei}, \citenamefont {Lebensohn}, \citenamefont {Suter},\ and\
  \citenamefont {Rollett}}]{ref:HEDM_4}%
  \BibitemOpen
  \bibfield  {author} {\bibinfo {author} {\bibfnamefont {R.}~\bibnamefont
  {Pokharel}}, \bibinfo {author} {\bibfnamefont {J.}~\bibnamefont {Lind}},
  \bibinfo {author} {\bibfnamefont {S.~F.}\ \bibnamefont {Li}}, \bibinfo
  {author} {\bibfnamefont {P.}~\bibnamefont {Kenesei}}, \bibinfo {author}
  {\bibfnamefont {R.~A.}\ \bibnamefont {Lebensohn}}, \bibinfo {author}
  {\bibfnamefont {R.~M.}\ \bibnamefont {Suter}}, \ and\ \bibinfo {author}
  {\bibfnamefont {A.~D.}\ \bibnamefont {Rollett}},\ }\bibfield  {title}
  {\enquote {\bibinfo {title} {In-situ observation of bulk 3d grain evolution
  during plastic deformation in polycrystalline cu},}\ }\href {\doibase
  https://doi.org/10.1016/j.ijplas.2014.10.013} {\bibfield  {journal} {\bibinfo
   {journal} {International Journal of Plasticity}\ }\textbf {\bibinfo {volume}
  {67}},\ \bibinfo {pages} {217--234} (\bibinfo {year} {2015})}\BibitemShut
  {NoStop}%
\bibitem [{\citenamefont {Pokharel}(2018)}]{ref:HEDM_1}%
  \BibitemOpen
  \bibfield  {author} {\bibinfo {author} {\bibfnamefont {R.}~\bibnamefont
  {Pokharel}},\ }\bibfield  {title} {\enquote {\bibinfo {title} {Overview of
  high-energy x-ray diffraction microscopy (hedm) for mesoscale material
  characterization in three-dimensions},}\ }in\ \href {\doibase
  https://doi.org/10.1007/978-3-319-99465-9} {\emph {\bibinfo {booktitle}
  {Materials Discovery and Design}}}\ (\bibinfo  {publisher} {Springer},\
  \bibinfo {year} {2018})\ pp.\ \bibinfo {pages} {167--201}\BibitemShut
  {NoStop}%
\bibitem [{\citenamefont {Pokharel}\ \emph {et~al.}(2014)\citenamefont
  {Pokharel}, \citenamefont {Lind}, \citenamefont {Kanjarla}, \citenamefont
  {Lebensohn}, \citenamefont {Li}, \citenamefont {Kenesei}, \citenamefont
  {Suter},\ and\ \citenamefont {Rollett}}]{ref:HEDM_2}%
  \BibitemOpen
  \bibfield  {author} {\bibinfo {author} {\bibfnamefont {R.}~\bibnamefont
  {Pokharel}}, \bibinfo {author} {\bibfnamefont {J.}~\bibnamefont {Lind}},
  \bibinfo {author} {\bibfnamefont {A.~K.}\ \bibnamefont {Kanjarla}}, \bibinfo
  {author} {\bibfnamefont {R.~A.}\ \bibnamefont {Lebensohn}}, \bibinfo {author}
  {\bibfnamefont {S.~F.}\ \bibnamefont {Li}}, \bibinfo {author} {\bibfnamefont
  {P.}~\bibnamefont {Kenesei}}, \bibinfo {author} {\bibfnamefont {R.~M.}\
  \bibnamefont {Suter}}, \ and\ \bibinfo {author} {\bibfnamefont {A.~D.}\
  \bibnamefont {Rollett}},\ }\bibfield  {title} {\enquote {\bibinfo {title}
  {Polycrystal plasticity: comparison between grain-scale observations of
  deformation and simulations},}\ }\href {\doibase
  https://doi.org/10.1146/annurev-conmatphys-031113-133846} {\bibfield
  {journal} {\bibinfo  {journal} {Annu. Rev. Condens. Matter Phys.}\ }\textbf
  {\bibinfo {volume} {5}},\ \bibinfo {pages} {317--346} (\bibinfo {year}
  {2014})}\BibitemShut {NoStop}%
\bibitem [{\citenamefont {Pandey}\ and\ \citenamefont
  {Pokharel}(2020)}]{ref:HEDM_3}%
  \BibitemOpen
  \bibfield  {author} {\bibinfo {author} {\bibfnamefont {A.}~\bibnamefont
  {Pandey}}\ and\ \bibinfo {author} {\bibfnamefont {R.}~\bibnamefont
  {Pokharel}},\ }\bibfield  {title} {\enquote {\bibinfo {title} {Machine
  learning enabled surrogate crystal plasticity model for spatially resolved 3d
  orientation evolution under uniaxial tension},}\ }\href
  {https://arxiv.org/pdf/2005.00951.pdf} {\bibfield  {journal} {\bibinfo
  {journal} {arXiv preprint arXiv:2005.00951}\ } (\bibinfo {year}
  {2020})}\BibitemShut {NoStop}%
\bibitem [{\citenamefont {Emma}\ \emph {et~al.}(2010)\citenamefont {Emma},
  \citenamefont {Akre}, \citenamefont {Arthur}, \citenamefont {Bionta},
  \citenamefont {Bostedt}, \citenamefont {Bozek}, \citenamefont {Brachmann},
  \citenamefont {Bucksbaum}, \citenamefont {Coffee}, \citenamefont {Decker}
  \emph {et~al.}}]{ref:LCLS}%
  \BibitemOpen
  \bibfield  {author} {\bibinfo {author} {\bibfnamefont {P.}~\bibnamefont
  {Emma}}, \bibinfo {author} {\bibfnamefont {R.}~\bibnamefont {Akre}}, \bibinfo
  {author} {\bibfnamefont {J.}~\bibnamefont {Arthur}}, \bibinfo {author}
  {\bibfnamefont {R.}~\bibnamefont {Bionta}}, \bibinfo {author} {\bibfnamefont
  {C.}~\bibnamefont {Bostedt}}, \bibinfo {author} {\bibfnamefont
  {J.}~\bibnamefont {Bozek}}, \bibinfo {author} {\bibfnamefont
  {A.}~\bibnamefont {Brachmann}}, \bibinfo {author} {\bibfnamefont
  {P.}~\bibnamefont {Bucksbaum}}, \bibinfo {author} {\bibfnamefont
  {R.}~\bibnamefont {Coffee}}, \bibinfo {author} {\bibfnamefont {F.-J.}\
  \bibnamefont {Decker}},  \emph {et~al.},\ }\bibfield  {title} {\enquote
  {\bibinfo {title} {First lasing and operation of an $\aa$ngstrom-wavelength
  free-electron laser},}\ }\href@noop {} {\bibfield  {journal} {\bibinfo
  {journal} {nature photonics}\ }\textbf {\bibinfo {volume} {4}},\ \bibinfo
  {pages} {641} (\bibinfo {year} {2010})}\BibitemShut {NoStop}%
\bibitem [{\citenamefont {Ju}\ \emph {et~al.}(2018)\citenamefont {Ju},
  \citenamefont {Highland}, \citenamefont {Thompson}, \citenamefont {Eastman},
  \citenamefont {Fuoss}, \citenamefont {Zhou}, \citenamefont {Dejus},\ and\
  \citenamefont {Stephenson}}]{ref:APSU}%
  \BibitemOpen
  \bibfield  {author} {\bibinfo {author} {\bibfnamefont {G.}~\bibnamefont
  {Ju}}, \bibinfo {author} {\bibfnamefont {M.~J.}\ \bibnamefont {Highland}},
  \bibinfo {author} {\bibfnamefont {C.}~\bibnamefont {Thompson}}, \bibinfo
  {author} {\bibfnamefont {J.~A.}\ \bibnamefont {Eastman}}, \bibinfo {author}
  {\bibfnamefont {P.~H.}\ \bibnamefont {Fuoss}}, \bibinfo {author}
  {\bibfnamefont {H.}~\bibnamefont {Zhou}}, \bibinfo {author} {\bibfnamefont
  {R.}~\bibnamefont {Dejus}}, \ and\ \bibinfo {author} {\bibfnamefont {G.~B.}\
  \bibnamefont {Stephenson}},\ }\bibfield  {title} {\enquote {\bibinfo {title}
  {Characterization of the x-ray coherence properties of an undulator beamline
  at the advanced photon source},}\ }\href@noop {} {\bibfield  {journal}
  {\bibinfo  {journal} {Journal of synchrotron radiation}\ }\textbf {\bibinfo
  {volume} {25}},\ \bibinfo {pages} {1036--1047} (\bibinfo {year}
  {2018})}\BibitemShut {NoStop}%
\bibitem [{\citenamefont {Maddali}\ \emph {et~al.}(2019)\citenamefont
  {Maddali}, \citenamefont {Allain}, \citenamefont {Cha}, \citenamefont
  {Harder}, \citenamefont {Park}, \citenamefont {Kenesei}, \citenamefont
  {Almer}, \citenamefont {Nashed},\ and\ \citenamefont
  {Hruszkewycz}}]{ref:BCDI:Sid1}%
  \BibitemOpen
  \bibfield  {author} {\bibinfo {author} {\bibfnamefont {S.}~\bibnamefont
  {Maddali}}, \bibinfo {author} {\bibfnamefont {M.}~\bibnamefont {Allain}},
  \bibinfo {author} {\bibfnamefont {W.}~\bibnamefont {Cha}}, \bibinfo {author}
  {\bibfnamefont {R.}~\bibnamefont {Harder}}, \bibinfo {author} {\bibfnamefont
  {J.-S.}\ \bibnamefont {Park}}, \bibinfo {author} {\bibfnamefont
  {P.}~\bibnamefont {Kenesei}}, \bibinfo {author} {\bibfnamefont
  {J.}~\bibnamefont {Almer}}, \bibinfo {author} {\bibfnamefont
  {Y.}~\bibnamefont {Nashed}}, \ and\ \bibinfo {author} {\bibfnamefont {S.~O.}\
  \bibnamefont {Hruszkewycz}},\ }\bibfield  {title} {\enquote {\bibinfo {title}
  {Phase retrieval for bragg coherent diffraction imaging at high x-ray
  energies},}\ }\href@noop {} {\bibfield  {journal} {\bibinfo  {journal}
  {Physical Review A}\ }\textbf {\bibinfo {volume} {99}},\ \bibinfo {pages}
  {053838} (\bibinfo {year} {2019})}\BibitemShut {NoStop}%
\bibitem [{\citenamefont {Brechb{\"u}hler}, \citenamefont {Gerig},\ and\
  \citenamefont {K{\"u}bler}(1995)}]{ref:Sph_1}%
  \BibitemOpen
  \bibfield  {author} {\bibinfo {author} {\bibfnamefont {C.}~\bibnamefont
  {Brechb{\"u}hler}}, \bibinfo {author} {\bibfnamefont {G.}~\bibnamefont
  {Gerig}}, \ and\ \bibinfo {author} {\bibfnamefont {O.}~\bibnamefont
  {K{\"u}bler}},\ }\bibfield  {title} {\enquote {\bibinfo {title}
  {Parametrization of closed surfaces for 3-d shape description},}\ }\href@noop
  {} {\bibfield  {journal} {\bibinfo  {journal} {Computer vision and image
  understanding}\ }\textbf {\bibinfo {volume} {61}},\ \bibinfo {pages}
  {154--170} (\bibinfo {year} {1995})}\BibitemShut {NoStop}%
\bibitem [{\citenamefont {Zhao}, \citenamefont {Wei},\ and\ \citenamefont
  {Wang}(2017)}]{ref:Sph_2}%
  \BibitemOpen
  \bibfield  {author} {\bibinfo {author} {\bibfnamefont {B.}~\bibnamefont
  {Zhao}}, \bibinfo {author} {\bibfnamefont {D.}~\bibnamefont {Wei}}, \ and\
  \bibinfo {author} {\bibfnamefont {J.}~\bibnamefont {Wang}},\ }\bibfield
  {title} {\enquote {\bibinfo {title} {Particle shape quantification using
  rotation-invariant spherical harmonic analysis},}\ }\href@noop {} {\bibfield
  {journal} {\bibinfo  {journal} {G{\'e}otechnique Letters}\ }\textbf {\bibinfo
  {volume} {7}},\ \bibinfo {pages} {190--196} (\bibinfo {year}
  {2017})}\BibitemShut {NoStop}%
\bibitem [{\citenamefont {Wei}, \citenamefont {Wang},\ and\ \citenamefont
  {Zhao}(2018)}]{ref:Sph_3}%
  \BibitemOpen
  \bibfield  {author} {\bibinfo {author} {\bibfnamefont {D.}~\bibnamefont
  {Wei}}, \bibinfo {author} {\bibfnamefont {J.}~\bibnamefont {Wang}}, \ and\
  \bibinfo {author} {\bibfnamefont {B.}~\bibnamefont {Zhao}},\ }\bibfield
  {title} {\enquote {\bibinfo {title} {A simple method for particle shape
  generation with spherical harmonics},}\ }\href {\doibase
  10.1016/j.powtec.2018.02.006} {\bibfield  {journal} {\bibinfo  {journal}
  {Powder technology}\ }\textbf {\bibinfo {volume} {330}},\ \bibinfo {pages}
  {284--291} (\bibinfo {year} {2018})}\BibitemShut {NoStop}%
\end{thebibliography}%

\end{document}